\documentclass[article]{jss}

\usepackage{orcidlink,thumbpdf,lmodern}

\usepackage{booktabs}
\usepackage{tabularx}
\usepackage{amsmath}
\usepackage{amssymb}
\usepackage{amsthm}
\usepackage{natbib}
\usepackage{tikz}
\usepackage{wrapfig}
\usepackage{arydshln}
\usepackage{array}
\usepackage{microtype}
\usepackage{todonotes}
\usepackage{lscape}
\usepackage{multirow}
\usepackage{makecell}
\usepackage{enumitem}
\usetikzlibrary{arrows.meta}
\usepackage{soul,color}
\usepackage[export]{adjustbox}
\usepackage{listings}
\usepackage{siunitx}   
\usepackage{caption}
\usepackage{graphicx} 
\usepackage{ragged2e}

\usepackage[autostyle, english = american]{csquotes}

\MakeOuterQuote{"}

\newcolumntype{M}[1]{>{\centering\let\newline\\\arraybackslash\hspace{0pt}}m{#1}}
\newcolumntype{L}[1]{>{\raggedright\let\newline\\\arraybackslash\hspace{0pt}}p{#1}}

\DeclareMathOperator*{\argmax}{arg\,max}

\author{
Markus Meierer\\University of Geneva \And
Patrick Bachmann\\ETH Zurich \And
Jeffrey N\"af\\University of Geneva \AND
Patrik Schilter\\University of Zurich \And
Ren\'e Algesheimer\\University of Zurich}
 \Plainauthor{Markus Meierer, Patrick Bachmann, Jeffrey N\"af, Patrik Schilter, Ren\'e Algesheimer}

\title{Estimating Individual Customer Lifetime Values with R: The \pkg{CLVTools} Package}
\Plaintitle{CLVTools}
\Shorttitle{\pkg{CLVTools}: Estimating Individual Customer Lifetime Values with \proglang{R}}

\Abstract{
Customer lifetime value (CLV) describes a customer's long-term economic value for a business. This metric is widely used in marketing, for example, to select customers for a marketing campaign. However, modeling CLV is challenging. When relying on customers' purchase histories, the input data is sparse. Additionally, given its long-term focus, prediction horizons are often longer than estimation periods. Probabilistic models are able to overcome these challenges and, thus, are a popular option among researchers and practitioners. The latter also appreciate their applicability for both small and big data as well as their robust predictive performance without any fine-tuning requirements. Their popularity is due to three characteristics: data parsimony, scalability, and predictive accuracy. The \proglang{R} package \pkg{CLVTools} provides an efficient and user-friendly implementation framework to apply key probabilistic models such as the Pareto/NBD and Gamma-Gamma model. Further, it provides access to the latest model extensions to include time-invariant and time-varying covariates, parameter regularization, and equality constraints. This article gives an overview of the fundamental ideas of these statistical models and illustrates their application to derive CLV predictions for existing and new customers. 
}

\Plainkeywords{customer lifetime value, probabilistic models, latent attrition models, customer spending models, customer analytics, R.}

\Address{
Markus Meierer\\
Geneva School of Economics and Management\\
University of Geneva\\
Boulevard du Pont-d'Arve 40 \\
1211, Geneva, Switzerland\\
E-mail: \email{markus.meierer@unige.ch}\\
URL: \url{http://www.unige.ch}
}

\begin{document}


\section[Introduction]{Introduction}

What is the future value of an individual customer to a business?  From a theoretical perspective, the concept of customer valuation and the use of customer lifetime value (CLV) as its key metric is well established \citep{Gupta2006}. CLV describes the long-term economic value of individual customers to a business. Building on predictive models of purchase behavior, the revenue of each customer is estimated in future periods. If available, these predictions are adjusted by respective cost projections. Then, following the idea of discounted cash flows, these predictions serve as a basis to calculate the net present value of each customer's future contributions to a business. The resulting metric informs and improves managerial decision-making. 

From a managerial perspective, the use of CLV in business practice is widespread. This is a consequence of marketers' increasing focus on customer-centricity. As the key to personalized offers, such a marketing strategy recognizes the importance of value-based segmentation and, consequently, the prioritization of customers \citep{Borle2008}. Having access to individual-level estimates of CLV allows firms to identify the most valuable customers. This helps to define how many resources to spend to maintain the relationship, e.g., by providing personalized services \citep{Gupta2006}. This also means that marketing spend on unprofitable customers can be minimized. Thus, marketers can increase their operational efficiency \citep[e.g.,][]{Reinartz2000}. In general, the regular assessment of CLV enables comprehensive benchmarking and optimization of customer acquisition, retention, and win-back campaigns. Moreover, the customer equity metric, i.e. summing up the CLV across all customers, is an important part of estimating the financial value of a firm \citep{McCarthy2018}. 

However, from a methodological point of view, predicting customer purchase behavior is not a straightforward task and has been an active area of research for decades \citep{Fader2014}. In particular, this applies to settings where customers do not have a formal contractual relationship with a business. Among others, this is the case for most companies in the retail or hospitality industry. In such non-contractual business settings, customers do not announce when they stop doing business with a firm. This significantly increases the complexity of the modeling task at hand. Such a setup lacks a well-defined dependent variable that could serve as input for a common regression-type model. To reliably predict customers' future behavior, a rigorous modeling approach would need to account for customers' unobserved attrition, transaction frequency, and average spending.

Providing a solution to this methodological challenge, practitioners and academics often turn to probabilistic models that capture customers' non-stationary purchase behavior through these three processes \citep[for a primer, see][]{Meierer2025b}. To model customers' attrition, transaction frequency, and spending process, the results of two separate models are usually combined. A first model is used to jointly model customers' attrition and transaction frequency. This allows us to determine how many transactions a customer will make until the business relationship is ended. Such models are known as latent attrition models. A second model captures customers' spending behavior, i.e., how much a customer spend on average per transaction. By combining the results of these two models it is possible to predict an individual customer's value to a business. In their simplest form, these models only need customers' transaction histories as input, i.e., a log file that includes the customer identifier, date, and value of the transaction. Despite this parsimoniousness, probabilistic approaches provide a remarkably robust assessment of customers' future value across industries. Their real-world applicability has been shown for office supply firms \citep{Schmittlein1994}, high-tech manufacturers \citep{Reinartz2003}, and retailers \citep{Abe2009, Fader2005c, Platzer2016}. In general, the probabilistic approach stands out from the alternatives. For one, it naturally accounts for customers' latent attrition. In addition, the underlying formal model of customer purchase behavior has shown to be a reliable foundation for making predictions, even when the forecasting horizon is much longer than the estimation period. Moreover, the parsimonious data requirements and scalability of the maximum likelihood estimation procedures provide advantages for real-world applications. Finally, recent research has proposed extensions to model exogenous and endogenous covariates, such as customer characteristics or seasonality \citep[e.g.,][]{Abe2009, Bachmann2021}. These extensions improve predictive accuracy and allow users to infer the incremental impact of marketing activities. By modeling time-varying covariates, such as exposure to promotions, firms can evaluate how specific interventions affect the future value of customers. 

\pkg{CLVTools} provides unified access to key probabilistic models to derive individual CLV predictions in a user-friendly way. First, the data preparation process for estimating these models is streamlined and required data transformations are abstracted away from the user. To generate the model inputs from transaction data, comparable packages require the user to execute a series of preparation steps scattered across various functions. Non-expert users lack understanding of the required sequence and variable definitions. In \pkg{CLVTools} all necessary data preparation is taken care of. Users only have to provide key definitions that characterize the transactional data of the specific business. Model-wise, basic models as well as their latest extensions, which enable users to accommodate time-invariant and -varying covariates, are implemented. Moreover, we allow users to impose additional specifications on the estimation process to infer correlations between key processes of customer behavior or to add $L_2$ regularization and equality constraints for covariate parameters. No other software package currently offers similar capabilities. The proposed \proglang{R} package \citep{r-core} thereby facilitates access to key probabilistic modeling approaches for practitioners and researchers.

The remainder of this paper is structured as follows. Section \ref{RelatedPackages} compares existing software implementations. Next, Section \ref{probmodels} provides an overview of probabilistic models for predicting customer purchase behavior. Thereby, we first introduce the Pareto/NBD model, which has been called the "gold standard" to jointly model customer attrition and purchase frequency \citep{Jerath2011}. We also outline recent modeling approaches to improve the flexibility in modeling customers' latent attrition. Furthermore, we discuss the most popular spending model, i.e., the Gamma-Gamma model \citep{Colombo1999, Fader2005b}. Combining the results of these two modeling tasks, the CLV for individual customers can be predicted. Bridging modeling theory and practice, Section \ref{guidance} then discusses modeling guidance from the literature and practice. Building on this, Section \ref{walkthrough} presents a case study from the retail industry to illustrate the application of \pkg{CLVTools} for predictive and inferential analyses. We conclude with directions for the future development of this \proglang{R} package.

\section[Related Software Packages]{Related Software Packages}\label{RelatedPackages}

While \pkg{CLVTools} provides unique features, it is not the first package to provide implementations of probabilistic models to predict individual CLV. The development of this \proglang{R} package was driven by two gaps in existing software. First, the lack of a unified interface makes model comparisons difficult. Second, the support for covariates is lacking for many model implementations. Beyond the obvious advantages, being able to model covariates also makes it possible to add further features which are highly sought after by applied scholars. This includes the possibility to add regularization and equality constraints on specific covariate parameters, as well as the ability to analyze the correlation between customers' attrition and transaction process. 

Table \ref{table:packagecomparison} provides a comparison of related software. 
Within the \proglang{R} ecosystem, the \pkg{BTYD} package \citep{BTYD} offers implementations of the most common latent attrition models and the \pkg{BTYDplus} package \citep{BTYDplus} provides implementations of additional latent attrition models, including one that is capable of considering time-invariant covariates. 
For \proglang{Python}, the package \pkg{Lifetimes} \citep{lifetimes} and its successor \pkg{PyMC-Marketing} \citep{PyMC-Marketing} provide basic functionality similar to the \pkg{BTYD} package. It is important to note that each of these packages supports different model types, and \pkg{CLVTools} does not aim to fully subsume them. For example, \pkg{BTYD} and \pkg{Lifetimes} implement the BG/BB model, a discrete-time analog of the continuous-time Pareto/NBD model.  \pkg{BTYDplus} offers extensions such as the Pareto/GGG model and a hierarchical Bayes version of the Pareto/NBD model. These specific models are not currently included in \pkg{CLVTools}, which currently focuses on continuous-time probabilistic models with closed-form marginal likelihoods.

\begin{table}[h!]
{\begin{tabular}{L{6.75cm} M{1.15cm} M{1.85cm} M{2.05cm} M{1.7cm}} 
    \hline
    \toprule 
    \textit{Features} &  \pkg{BTYD} (\proglang{R}) &   \pkg{BTYDplus} (\proglang{R}) &  \pkg{CLVTools}\phantom{**} (\proglang{R})\phantom{**} & \pkg{Lifetimes*} (\proglang{Python}) \\
    \midrule 
    Usage of classes \& generic methods & no & no & yes\phantom{**} & yes \\
    Time-invariant covariates  &  no & yes & yes\phantom{**} & no \\
    Time-varying covariates   & no & no & yes** & no \\
    Correlation between processes  & no & no & yes** & no \\
    Regularization for covariate param.  & no & no & yes\phantom{**} & no \\
    Equality constraints for covariate param. & no & no & yes\phantom{**} & no \\
    Abstraction layer for data preparation  & no & no & yes\phantom{**} & no \\
    \bottomrule
    \multicolumn{5}{p{.95\textwidth}}{\scriptsize Notes: * Also applicable to its successor \pkg{PyMC-Marketing}. ** At the time of writing, derivations  only exist for the Pareto/NBD model and its variants. Thus, these features are only implemented for these models.} 
    \end{tabular}}
\caption{Software packages implementing probabilistic models for CLV modeling\label{table:packagecomparison}}
\end{table}

\pkg{CLVTools} builds on key lessons from surveying the functionality and workflow of these three packages. We provide all functionalities in a class-based (S4) framework and work with the canonical generic methods that \proglang{R} users are familiar with \citep{Schilter2025}. It is also optimized to handle large datasets with millions of purchase records.

It should not be ignored that \pkg{CLVTools} is made possible by key contributions offered in the following packages: \pkg{data.table} \citep{data.table}, \pkg{ggplot2} \citep{ggplot2}, \pkg{lubridate} \citep{lubridate}, \pkg{Matrix} \citep{Matrix}, \pkg{MASS} \citep{MASS}, \pkg{optimx} \citep{optimx1, optimx2}, \pkg{Rcpp} \citep{rcpp1}, \pkg{RcppArmadillo} \citep{rcpparamdillo}, \pkg{RcppGSL} \citep{rcppgsl}, \pkg{stats} \citep{r-core}, and \pkg{utils} \citep{r-core}.

\section[Probabilistic modeling of customer purchase behavior]{Probabilistic modeling of customer purchase behavior}\label{probmodels}

Probabilistic models, also called stochastic or probability models, are an important part of the marketing research literature, particularly for modeling customer purchase behavior in non-contractual business settings. Since their first use to predict purchase patterns by \citet{Magee1953}, significant progress has been made.  \citet{Fader2014} provide a detailed overview of their use and evolution throughout the following decades.  

Compared to alternative approaches, probabilistic modeling of purchase behavior has several advantages \citep{Fader2006}: First, probabilistic models do not require observed transaction data to be split into two periods to create a dependent variable. As they explicitly model the data generating process, there is no need for a separate holdout set that defines the outcome measure of interest, e.g., customers' next year revenue. Thus, all data can be used for model estimation \citep{Fader2009}. This is in contrast to many supervised learning techniques. Second, covariates can be included, but they are not a requirement. Indeed, the parsimoniousness of the models' data requirements has been a key reason for their widespread real-world adoption. The basic probabilistic modeling approach only requires the transactional records of a business consisting of three data columns, i.e., the date and monetary value of the transaction as well as a customer identifier. Third, the underlying formal model of customer purchase behavior has been shown to be a reliable foundation for making predictions. This even applies when the forecasting horizon is much longer than the estimation period. Fourth, the scalability of many probabilistic models is high. Their application to millions of customers does not require specialized hardware such as graphics processing units. Fifth, recent advancements have introduced inferential analyses capabilities by allowing the inclusion of both time-invariant and -varying covariates. This enables comprehensive explanatory analyses and, thus, extends the applicability of the models beyond the predictive domain. This interpretability further distinguishes these models from machine learning approaches. 

In the following section, we give an overview of the general modeling structure for inferring customer lifetime values for a non-contractual business and thereby, distinguish two key modeling tasks. Next, we explain the mathematical intuition of a representative probabilistic model for each modeling task. Namely, our discussion focuses on the Pareto/NBD model to account for customers' attrition and transaction process, as well as the Gamma-Gamma model to account for customers' spending process. We also detail respective model extensions for incorporating covariates as well as advanced techniques such as parameter regularization. While further models are implemented in \pkg{CLVTools}, their general logic is analogous. The derivations for the Beta-Geometric/NBD (BG/NBD) model are found in \cite{Fader2005c} and for the Gamma-Gompertz/NBD (GGomp/NBD) model in \citet{Bemmaor2012}. Note that we use the term CLV to denote the future contributions of customers, i.e., projections that build on their observed past transaction behavior. In the literature, this is sometimes referred to as \emph{residual} customer lifetime value.

\subsection{General modeling structure } 

Probabilistic approaches focus on three processes to derive customers' future purchase behavior and thus, individual CLVs. These are customers'
\begin{itemize}[topsep=0em]
\setlength\itemsep{0em}
\item attrition process, i.e., "how long will a customer be active?", 
\item transaction process, i.e., "how many times will a customer purchase?", and 
\item spending process, i.e., "how much will a customer spend on every purchase?".
\end{itemize}

The first two processes are usually estimated jointly, modeling the unobserved attrition process of customers in combination with the individual transactional frequency. It is often referred to as a "buy-till-you-die" model or latent attrition model. Thereby, it is assumed that a customer buys according to a heterogeneous stationary process and then “dies” at an unobserved, as-if random point in time. This structure has proven to be very effective in many empirical settings \citep{Fader2014}. The third process is usually assumed to be independent and is modeled separately. It infers the average spending per purchase acknowledging customer heterogeneity. 

Estimating these three processes with two separate models results in a customer-specific prediction for two metrics: (1) discounted expected transactions and (2) the average monetary value of each transaction. A customer's future value to a business is then calculated as the product of these two metrics. In the following, let $\mathcal{F}$ denote the information we observe from a generic customer. For instance, for the Pareto/NBD model introduced below, $\mathcal{F}=\{x,t_x,T \}$, the number of transactions, timing of the last observed transaction, and the end of the observation period respectively. When predicting future customer purchase behavior, we know neither the lifetime of an individual customer nor the timing and nature of their purchase while they are "alive" as a customer \citep{Ascarza2017}. Considering these quantities as random variables, our objective is to derive the conditional expectation of a customer's future value to a business $E(CLV \mid \mathcal{F})$, her expected number of transactions in the future $E(X(T,T+t)\mid \mathcal{F})$, and the spending per transaction $E(m\mid \mathcal{F})$. Following \citet{Rosset2003}, we state that

\begin{equation}\label{eq:CLV}
E(C L V\mid \mathcal{F})=\int_T^{\infty} E[V(t)\mid \mathcal{F}] S(t\mid \mathcal{F}) d(t) d t
\end{equation}

with $t=T$ representing the end of the observation period of the customer.  
$E[V(t)\mid \mathcal{F}]$ is the expected cash flow from a customer at time $t$, assuming that they are alive at that time. $S(t\mid \mathcal{F})$ is the probability that a customer has remained alive for at least time $t$, and $d(t)$ is a discount factor that reflects the present value of money received at time $t$.

Probabilistic modeling of customer value builds on this foundation. Again, an underlying assumption is that the monetary value of a particular transaction is independent of the transaction process, i.e., how many transactions a customer will make in the future. Thus, the cash flow per transaction can be factored out of the calculation. While this assumption can be relaxed, probabilistic modeling of CLV usually focuses on both predicting the sequence of future transactions and predicting the average monetary value of customers' transactions, i.e., $E(m\mid \mathcal{F})$. The former includes a discounting of future transactions to their present value. The resulting metric is called discounted expected residual transactions ($DERT$):
\begin{align}
DERT=\int_T^{\infty} \lambda(t) S(t\mid \mathcal{F}) d(t) dt, 
\end{align}
where $\lambda(t)$ is the purchase rate \citep{Fader2006,Bachmann2021}. This leads to 
\begin{align}
E(CLV\mid \mathcal{F})       
= E(m\mid \mathcal{F}) DERT. 
\end{align}

Expressions for DERT, $E(m\mid \mathcal{F})$ as well as $E(X(T,T+t)\mid \mathcal{F})$ are available for all variations of the Pareto/NBD model. In the following $E(X(T,T+t)\mid \mathcal{F})$ will be referred to as conditional expected transactions (CET). Note that multiplying DERT and $E(m\mid \mathcal{F})$  makes the assumption that spending is stationary. The latter might require a transformation such as inflation adjustment, as mentioned in \ref{guidance}. Forecasting time-varying spending and adjusting CLV estimates accordingly is a subject of ongoing research.

The decision on which models to use for these two modeling tasks is guided by theoretical and practical considerations (see Section \ref{guidance}). Whether for latent attrition or customer spending models, the difference between probabilistic model alternatives lies mainly in the distributional assumptions of the underlying processes. However, model extensions, which add the ability to include time-invariant and -varying covariates, have only been proposed recently and are only available for some models.

\subsection{Modeling customers' latent attrition: The Pareto/NBD model}

In the following, we provide the mathematical intuition behind the Pareto/NBD model. For technical details, see \citet{Schmittlein1987} and \citet{Fader2005}. As with other latent attrition models, the Pareto/NBD consists of two parts. The first process focuses on modeling customers' unobserved attrition. A second process considers customer purchase frequency given the assumption that they are still active. Customers "live" and therefore buy for an unknown period of time until they "die", i.e., become inactive.

Considering customer attrition, a customer's unobserved lifetime of length $\omega$ is assumed to be exponentially distributed with rate $\mu$ and density
\begin{equation}\label{eq:standardlifetime}
f(\omega | \mu)= \mu e^{- \mu \omega}.
\end{equation}
$\mu$ is the mean attrition rate and is itself assumed to follow a Gamma distribution with shape parameter $s$ and scale parameter $\beta$ to account for the cross-sectional heterogeneity 
\begin{equation}\label{eq:heterogenitydistlifetime}
g(\mu)=\frac{\beta^s \mu^{s-1} e^{-\mu \beta}}{\Gamma(s)}.
\end{equation}

In the original paper, \citet{Schmittlein1987} refer to $\mu$ as "average death rate" (p. 11).

Combining the Exponential distribution of $\mu$ with the Gamma distribution in (\ref{eq:heterogenitydistlifetime}) results in a Pareto distribution of the second kind. The density of the lifetime of a random customer is:
\begin{align}\label{eq:standardtranswithhetero}
f(\omega) & =\int_{0}^{\infty} f(\omega | \mu) g(\mu) d\mu 
= \frac{s}{\beta} 
\left(\frac{\beta}{\beta + \omega}\right)^{s+1}. 
\end{align}

Looking at the transaction process, the Pareto/NBD model assumes that transactions follow a Poisson process with rate $\lambda$ for any given customer. Thus, the probability of observing $x$ transactions in the time interval $(0,t]$ follows a Poisson distribution with rate $\lambda t$, i.e.,
\begin{equation}\label{eq:Poisson}
P(X(t)=x| \lambda, t < \omega)=\frac{(\lambda t)^x e^{-\lambda t }}{x !}, \ \ x=0,1,2,\ldots,
\end{equation}

where $\lambda$ is interpreted as the mean purchase rate of a given customer. Accounting for heterogeneity, $\lambda$ follows a Gamma distribution with shape parameter $r$ and scale parameter $\alpha$:

\begin{equation}\label{eq:heterogenitydisttrans}
g(\lambda)=\frac{\alpha^r \lambda^{r-1} e^{-\lambda \alpha}}{\Gamma(r)}.
\end{equation}

This combination of (\ref{eq:Poisson}) and (\ref{eq:heterogenitydisttrans}) results in a negative binomial distribution (NBD). Thus, the number of transactions made by a given customer while "alive" ($t < \omega$) is given by: 
\begin{align}\label{eq:standardtrans}
P(X(t)=x \mid t < \omega) & =\int_{0}^{\infty} P(X(t)=x| \lambda,t < \omega ) g(\lambda ) d \lambda 
= \frac{\Gamma(r + x)}{\Gamma(r) x !} 
\left(\frac{\alpha}{\alpha + t} \right)^r 
\left(\frac{t}{\alpha + t} \right)^x.
\end{align}

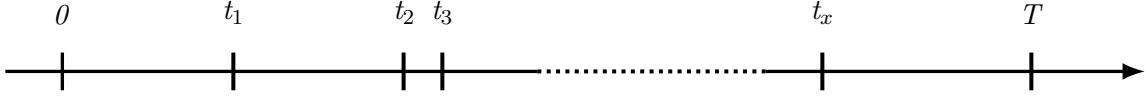
\begin{figure}[h!]
\centering
\vspace{0.1cm}
    
\begin{tikzpicture}
    \draw[-, line width=1.25pt] (0,0) -- (7,0);
    \draw[dotted, line width=1.5pt] (7,0) -- (10,0);
    \draw[-{Latex[length=3mm]}, line width=1.25pt] (10,0) -- (15,0);
        
    \draw[black, line width=1.25pt] (0.75,-0.25) -- +(0,0.5cm);
    \node[above] at (0.75,0.5cm) {\textit{0}};
        
    \draw[black, line width=1.5pt] (3,-0.25) -- +(0,0.5cm);
    \node[above] at (3,0.5cm) {\textit{t$_1$}};
        
    \draw[black, line width=1.5pt] (5.24,-0.25) -- +(0,0.5cm);
    \node[above] at (5.25,0.5cm) {\textit{t$_2$}};

    \draw[black, line width=1.5pt] (5.75,-0.25) -- +(0,0.5cm);
    \node[above] at (5.75,0.5cm) {\textit{t$_3$}};
        
    \draw[black, line width=1.5pt] (10.75,-0.25) -- +(0,0.5cm);
    \node[above] at (10.75,0.5cm) {\textit{t$_x$}};
        
    \draw[black, line width=1.5pt] (13.5,-0.25) -- +(0,0.5cm);
    \node[above] at (13.5,0.5cm) {\textit{T}};
        
\end{tikzpicture}
    
\vspace{0cm}
\caption{Schematic transaction history for a prototypical customer}
\label{timeline}
\end{figure}

Combined in a single likelihood function, these two processes formalize the generative process of this probabilistic model of purchase behavior. As illustrated in Figure \ref{timeline}, suppose that a given customer makes $x$ purchases during the observation period $(0, T]$ at times $t_1, t_2, ..., t_x$. Note the uneven timings of transactions that characterize their stochastic nature.  

A customer is still alive at $T$ (i.e., $\omega > T $) or has churned between the last transaction $t_x$ and the observation end $T$ (i.e., $\omega \in (t_x, T]$ ). The likelihood function expresses these two possibilities. If a customer is alive until $T$, the "likelihood function is simply the product of the (inter-transaction-time) exponential density functions and the associated survivor function" \citep[p. 5]{Fader2005} and can be expressed as in Eq. (\ref{Eq:pnbd_ll_untilT}). In cases where a customer became inactive at some time $\omega$ in the interval $(t_x, T]$, the individual-level likelihood function is as in Eq. (\ref{Eq:pnbd_ll_beforeT}). The expressions are very similar and differ only in their conditioning on whether the customer "dies" before or after $T$:    
\begin{align}
    L\left(\lambda \mid t_{1}, \ldots, t_{x}, T, \omega>T\right)  
      & =\lambda^{x} e^{-\lambda T}  \label{Eq:pnbd_ll_untilT}       \\
    L\left(\lambda \mid t_{1}, \ldots, t_{x}, T, \text { inactive at } \omega \in\left(t_{x}, T\right]\right)  
      & =\lambda^{x} e^{-\lambda \omega}  \label{Eq:pnbd_ll_beforeT} 
\end{align}
    
For an individual customer with a purchase summary $(x, t_x, T)$, the likelihood function is the combination of these expressions, weighted by their respective probabilities. Note that the exact timing of transactions, except the last one at $t_x$, becomes irrelevant in this model due to the memoryless property of the exponential distribution: 
\begin{align}
    L\left(\lambda, \mu \mid x, t_{x}, T\right) & = L(\lambda \mid x, T, \omega>T) P(\omega>T \mid \mu)  \nonumber                                                                                             \\
                                                & \phantom{=} +\int_{t_{x}}^{T} L\left(\lambda \mid x, T, \text { inactive at } \omega \in\left(t_{x}, T\right]\right) f(\omega \mid \mu) d \omega   \nonumber \\
                                                & = \frac{\lambda^{x} \mu}{\lambda+\mu} e^{-(\lambda+\mu) t_{x}}+\frac{\lambda^{x+1}}{\lambda+\mu} e^{-(\lambda+\mu) T} \label{Eq:indLL}                       
\end{align}
    
Customers' latent characteristics $\lambda$ and $\mu$ are then marginalized over the two Gamma distributions. The result is a rather intricate closed-form expression from which values for  $r$, $\alpha$, $s$, and $\beta$ are estimated using a maximum likelihood approach (see Appendix \ref{app:pnbd_ll}). Parameters $r$ and $\alpha$ are the shape and scale parameters of the Gamma distribution of the transaction process, and $s$ and $\beta$ are the shape and scale parameters of the Gamma distribution of the attrition process, respectively. Consequently, $r/\alpha$ and $s/\beta$ represent the purchase rate and the dropout rate of an average customer, respectively. Given transactional records for a total of $N$ customers, these parameters can be estimated by maximizing the log-likelihood
\begin{align}
    \sum_{i=1}^N \log L(r, \alpha, s, \beta | x_i, t_{x_i}, T_i). 
\end{align}
    
The likelihood reflects the underlying generative process of the Pareto/NBD model. Also, it highlights its data parsimony; only $x$, $t_{x}$, and $T$ are required as input. However, the latent parameters $\lambda$ and $\mu$ are assumed to be independent but this may be relaxed (see Section \ref{advanced-modeling}). For the conditional expectation and managerial expressions, see \citet{Fader2005}.
    
In sum, the Pareto/NBD model is a parsimonious representation of customers' attrition and transaction process. Yet, its standard form has shortcomings, e.g., covariates cannot be added.

\subsection{The Pareto/NBD model with time-invariant and -varying covariates} \label{timevariant}

Thus far, heterogeneity has been solely modeled by Gamma distributions. However, observable information, such as customer demographics, is often available. These covariates may help to explain part of the heterogeneity among customers and, therefore, increase the predictive accuracy of the model. In addition, we can rely on their parameter estimates for inference, i.e., to identify and quantify the effects of time-invariant covariates on the transaction and attrition process. The extension for time-invariant covariates was discussed in \citet{Fader2007c}. More generally, in many real-world applications, customer transaction and attrition behavior may be influenced by covariates that \emph{vary over time}. Consequently, the timing of a purchase and the corresponding value of the covariate at that time become relevant. Time-varying covariates can affect customers on the aggregated level as well as on the individual level. In the first case, all customers are affected simultaneously; in the latter case, a covariate is only relevant to a particular customer. In general, the literature identifies two groups of time-varying covariates \citep{Platzer2016}: (1) seasonality in purchase patterns and (2) marketing activities. Seasonality in purchase patterns is observed in many industries. These patterns can be caused on the aggregated level by holidays or on the individual level by recurring personal events such as birthdays or paydays. Marketing activities also significantly influence customer behavior. Firms target their customers on the aggregated level through mass marketing and on the individual level through direct marketing campaigns. \citet{Bachmann2021} address this issue and introduce time-varying covariates to the Pareto/NBD model. We now explain the key ideas behind this extension. For a comprehensive discussion of the technical details, we refer to the original paper.
    
\begin{figure}[h!]
    \centering
    \vspace{0cm}
        
    \begin{tikzpicture}
        \draw[-, line width=1.25pt] (0,0) -- (7,0);
        \draw[dotted, line width=1.5pt] (7,0) -- (10,0);
        \draw[-{Latex[length=3mm]}, line width=1.25pt] (10,0) -- (15,0);
            
        \draw[black, line width=1.25pt] (0.75,-0.25) -- +(0,0.5cm);
        \node[above] at (0.75,0.5cm) {\textit{0}};
            
        \draw[black, line width=1.5pt] (3,-0.25) -- +(0,0.5cm);
        \node[above] at (3,0.5cm) {\textit{t$_1$}};
            
        \draw[black, line width=1.5pt] (5.25,-0.25) -- +(0,0.5cm);
        \node[above] at (5.25,0.5cm) {\textit{t$_2$}};

        \draw[black, line width=1.5pt] (5.75,-0.25) -- +(0,0.5cm);
        \node[above] at (5.75,0.5cm) {\textit{t$_3$}};
            
        \draw[black, line width=1.5pt] (10.75,-0.25) -- +(0,0.5cm);
        \node[above] at (10.75,0.5cm) {\textit{t$_x$}};
            
        \draw[black, line width=1.5pt] (13.5,-0.25) -- +(0,0.5cm);
        \node[above] at (13.5,0.5cm) {\textit{T}};

        \draw[dotted, thick] (0.5,0.2) -- (0.5,-0.7);
        \draw[dotted, thick] (2.5,0.2) -- (2.5,-0.7);
        \draw[dotted, thick] (4.5,0.2) -- (4.5,-0.7);
        \draw[dotted, thick] (6.5,0.2) -- (6.5,-0.7);
        \draw[dotted, thick] (10.5,0.2) -- (10.5,-0.7);
        \draw[dotted, thick] (12.5,0.2) -- (12.5,-0.7);

        \node[below] at (1.5,-0.3) {$\mathbf{x}_1^P$, $\mathbf{x}_1^A$};
        \node[below] at (3.5,-0.3) {$\mathbf{x}_2^P$, $\mathbf{x}_2^A$};
        \node[below] at (5.5,-0.3) {$\mathbf{x}_{3}^P$, $\mathbf{x}_3^A$};
        \node[below] at (11.5,-0.3) {$\mathbf{x}_{k_T-1}^P$, $\mathbf{x}_{k_T-1}^A$};
        \node[below] at (13.5,-0.3) {$\mathbf{x}_{k_T}^P$, $\mathbf{x}_{k_T}^A$};
            
    \end{tikzpicture}
            
    \caption{Schematic transaction history of a prototypical customer with time-varying covariates}
    \label{timeline:covariates}
\end{figure}
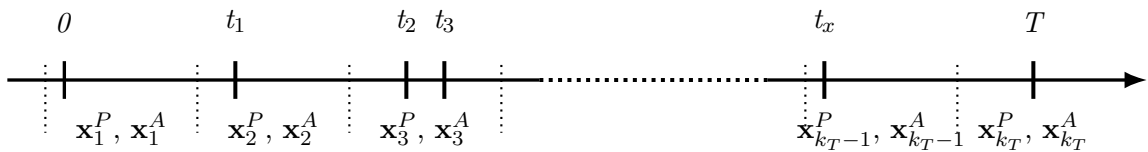

The foundation of time-varying covariates are the functions $\mathbf{x}^P(s)$, $\mathbf{x}^A(s)$ which map a time point $s$ to a vector of covariates at time point $s$. Note that these functions are individual-level quantities and can
generally vary between customers, although we omit the subscript $i$ for readability reasons. Furthermore, it is key to understand that these functions are parameterized by assuming equidistant time intervals (e.g., one week) during which the covariates are assumed to be constant. That is, $\mathbf{x}^P(s)$, $\mathbf{x}^A(s)$ are piecewise constant functions that return the covariates $\mathbf{x}^P_{k}, \mathbf{x}^A_{k}$ of the interval $k$ in which $s$ lies, as in Figure \ref{timeline:covariates}.

The length of these intervals can be arbitrarily small and the discretization does not affect the continuous nature of the transaction and attrition process. The covariates for the transaction process ($\mathbf{x}^P$) and attrition process ($\mathbf{x}^A$) may differ. We then obtain a time-varying mean purchase rate $\lambda(t)$ and customer attrition rate $\mu(t)$, where
\[
    \lambda(t) = \lambda_0 \exp(\boldsymbol{\gamma}_{purch}'\mathbf{x}^P(t)), \ \  \mu(t) = \mu_0 \exp(\boldsymbol{\gamma}_{attr}'\mathbf{x}^A(t)),
\] 
with $\lambda_0, \mu_0$ following the Gamma distributions in \eqref{eq:heterogenitydisttrans} and \eqref{eq:heterogenitydistlifetime} respectively.

For the transaction process, the probability of purchasing $x$ times is now dependent on the covariates in the relevant time period. The probability of $x$ transactions in the interval $(s_1, s_2]$, with two time points $s_1 < s_2$, is given as
\begin{equation}\label{eq:timevaryingtrans}
    P(X(s_1,s_2)=x \mid \lambda(t))=\frac{[\Lambda(s_1, s_2)]^x}{x!} \exp[-\Lambda(s_1, s_2)],
\end{equation}
with
\begin{equation}	\label{thetapT}
        \Lambda(s_1, s_2)= \int_{s_1}^{s_2} \lambda(s) d s.
\end{equation}
In contrast to assumptions of models without time-varying covariates, the times between purchases are no longer independently and identically distributed. In particular, the distribution of the time between purchases $j-1$ and $j$ depends on the time of the last purchase $j-1$. This determines which covariate time interval affects this particular mean purchase rate.
    
For the attrition process, the density is dependent on the covariates across all time points, i.e.,
\begin{equation}\label{eq:covdynamiclifetime}
    f(\omega | \mu(t))= \mu(t) \exp[-M(\omega)],
\end{equation}
with
\begin{equation} \label{thetapL}
    M(\omega)= \int_{0}^{\omega} \mu(s) d s.
\end{equation}
    
Combining both processes results in  
$L(\alpha, r , \beta, s, \boldsymbol{\gamma}_{purch}, \boldsymbol{\gamma}_{attr}|\mathbf{X}^P,\mathbf{X}^A, x , \mathbf{t}, T)$ This closed-form likelihood contains the covariates for every time interval $k$,
\begin{align*}
    \mathbf{X}^P                                  
    =                                             
    ( \mathbf{x}_1^P, \ldots,  \mathbf{x}_{k_T}^P 
    ) \text{  and  }                              
    \mathbf{X}^A                                  
    =                                             
    ( \mathbf{x}_1^A, \ldots,  \mathbf{x}_{k_T}^A 
    ).                                            
\end{align*}
Instead of only considering the recency of the last transaction as in the standard model, this extension considers the exact date of all transactions $\mathbf{t}=\left(t_1,\ldots, t_x \right)$. This likelihood only contains a sum of expressions similar to the likelihood of the standard model, each being calculated for a certain time interval (see \citet[Appendix A.1]{Bachmann2021}). 
    
The standard model and the extension for time-invariant covariates are nested within this model.  With covariate effects set to zero, we arrive at the standard model. When including time-invariant and -varying covariates, the former maintain the same parameter value over all time intervals. $\alpha$, $r$, $\beta$, $s$, $\boldsymbol{\gamma}_{purch}$ and $\boldsymbol{\gamma}_{attr}$ are estimated using maximum likelihood. For the conditional expectation and managerial expressions, see \citet{Bachmann2021}.

\subsection{Advanced modeling techniques for latent attrition models}\label{advanced-modeling}

To adapt latent attrition models to the specific application context, several advanced modeling techniques can be leveraged. These include (1) correlating customers' attrition and transaction process, (2) regularizing covariate parameter estimates $\boldsymbol{\gamma}$, and (3) adding equality constraints for the covariate parameter estimates $\boldsymbol{\gamma}$.  

To begin with, we look at the correlation of customers' attrition and transaction processes. All models discussed so far assume independence between mean transaction and  attrition rates. This may not be a realistic assumption. Thus, we use Sarmanov distributions to correlate the mean transaction and mean attrition rates \citep{Park2004}. For alternatives see, e.g., \citet[][]{Glady2015}. We will refer to the example of the standard Pareto/NBD model but the approach is analogously applicable for related latent attrition models.
    
The Sarmanov distribution family provides a straightforward way to introduce dependence between the mean purchase rate $\lambda$ and the mean attrition rate $\mu$. With the Sarmanov approach, the joint density of $(\lambda,\mu)$ can be expressed as a combination of multiple univariate densities that describe only one random variable. That is, the joint density can be written as the sum of the products of the univariate densities of $\lambda$ and $\mu$:
\begin{align}\label{cooltrick}
    g(\lambda,\mu|\alpha,r, \beta, s, m)= & g(\lambda|r,\alpha)g(\mu|s,\beta) + m \left(\frac{\alpha}{1 + \alpha} \right)^r \left(\frac{\beta}{1 + \beta} \right)^{s}\notag \\
    & \cdot \big[ (g(\lambda|r,\alpha+1) - g(\lambda|r,\alpha)) (g(\mu|s,\beta+1) - g(\mu|s,\beta)) \big]. 
\end{align}

Intuitively, this distribution combines the independent distribution $g(\lambda|r,\alpha)g(\mu|s,\beta)$ with a term that induces dependence. When $m$ is positive, the formula increases probability density in regions where both variables are similarly positioned (both high or both low) and decreases it elsewhere. This is because $g(\lambda|r,\alpha+1)$ is smaller than $g(\lambda|r,\alpha)$ for ``high'' values of $\lambda$, making this difference negative. Similarly, $g(\mu|s,\beta+1)-g(\mu|s,\beta)$ is negative for $\mu$ ``high''. Their product becomes positive in these regions, creating positive dependence between $\lambda$ and $\mu$. The opposite occurs when $m$ is negative, resulting in negative dependence.

The correlation of the two processes of the Pareto/NBD model is thereby given as
\begin{equation} \label{eq:corrkoefficient}
    p_{m}=m \frac{\sqrt{r}}{1+\alpha} \left( \frac{\alpha}{1+\alpha} \right)^r \frac{\sqrt{s}}{1+\beta} \left( \frac{\beta}{1+\beta} \right)^s.
\end{equation}
As a consequence of Eq. (\ref{cooltrick}), the new individual likelihoods of customers can be expressed as the sum of those likelihoods without correlation. Given the likelihood without correlation  
\[
    L(\alpha,r,\beta,s, m| x,t,T)=\tilde{L}(\alpha,\beta),
\]
it follows that the likelihood with correlation can be expressed as
\begin{equation}
    \label{eq:corrll}
    \begin{split}
        &L(\alpha,r,\beta,s, m|x,t,T)= \tilde{L}(\alpha,\beta) + m \left(\frac{\alpha}{1 + \alpha} \right)^r \left(\frac{\beta}{1 + \beta} \right)^{s} \\
        &\times \big[ \tilde{L}(\alpha + 1,\beta + 1)-  
            \tilde{L}(\alpha+1,\beta) - \tilde{L}(\alpha,\beta+1)
        + \tilde{L}(\alpha,\beta)\big].
    \end{split}
\end{equation}
    
The new likelihood contains the additional parameter $m$ that captures the correlation between the underlying transaction and  attrition process. Note that this coefficient must not be directly interpreted as a correlation coefficient, but Eq. (\ref{eq:corrkoefficient}) can be used to do so.

Another modeling technique that can help to estimate latent attrition models with many covariates, is regularization. In these cases, regularization may be used to avoid overfitting and numerical instability. It is applied to the parameter vectors $\boldsymbol{\gamma}_{purch}$ and $\boldsymbol{\gamma}_{attr}$ for the covariates. To add regularization, we put a Gaussian prior on $\boldsymbol{\gamma}_{purch}$ and $\boldsymbol{\gamma}_{attr}$ such that
\[
    \boldsymbol{\gamma}_{purch} \sim N(0, 1/\lambda^{reg}_{1} I), \boldsymbol{\gamma}_{attr} \sim N(0, 1/\lambda^{reg}_{2} I)
\]
for some hyperparameters $\lambda^{reg}_{1}$ and $\lambda^{reg}_{2}$, and $I$ is the identity matrix. Note that the symbol $\lambda$ used here for the regularization strength ($\lambda^{reg}_1$, $\lambda^{reg}_2$) is conceptually distinct from the $\lambda$ parameter in the generative BTYD models, which denotes the individual purchase rate. 
    
In this case, the maximum a posteriori optimization problem is given by

\begin{align}
    \argmax_{\substack{\alpha,r,\beta,s \\ \boldsymbol{\gamma}_{purch},\boldsymbol{\gamma}_{attr}}}
      & \sum_{i=1}^{n} L(\alpha,r,\beta,s,\boldsymbol{\gamma}_{purch},\boldsymbol{\gamma}_{attr}| \mathbf{X}^P_i, \mathbf{X}^A_i,x_i,\mathbf{t}_i,T_i) - \lambda^{reg}_{1} \| \boldsymbol{\gamma}_{purch}\|^2 - \lambda^{reg}_{2} \| \boldsymbol{\gamma}_{attr}\|^2, 
    \label{regularization}
\end{align}
 
which corresponds to the L2-norm regularization of the likelihood.

A final modeling technique that focuses on extending the possibilities for inferential analyses is equality constraints for covariate parameters of latent attrition models. These equality constraints make it possible for the optimizer to always use the same covariate parameter estimates for the transaction and attrition process. The implementation is straightforward: 
\begin{align}
    \boldsymbol{\gamma}_{purch}\equiv\boldsymbol{\gamma}_{attr}, 
\end{align}
for $\lambda_0 \exp(\boldsymbol{\gamma}_{purch}'\mathbf{x}^P)$ and $\mu_0 \exp(\boldsymbol{\gamma}_{attr}'\mathbf{x}^A)$ with $\mathbf{x}^P= \mathbf{x}^A$. 
 
This feature enables researchers to test hypotheses. The equality constraints enable to test for any covariate whether or not there is a difference between the impact of a covariate on the two processes. The model that constrains both parameter estimates serves as a benchmark model to be able to do a likelihood ratio test and thus to determine formally if these parameter estimates are indeed different. For example, relying on a likelihood ratio test, we could assess whether direct marketing activities significantly increase the number of transactions while simultaneously leading to significantly higher churn.

\subsection{Modeling customers' spending process: The Gamma-Gamma model}\label{spending}
Latent attrition models usually only focus on customers' attrition and transaction process. However, to derive a monetary value such as CLV, customers' spending behavior must also be considered. Here, the Gamma-Gamma model is a popular choice. For a discussion of technical details beyond the following overview, see \citet{Colombo1999} and \citet{Fader2005b}.
    
The true mean spending rate $\zeta$ of a customer is not observed. Therefore, we base each individual prediction on a customer's past average spending value $\bar{z}$, where
\begin{align}
    \label{z_avg}                    
    \bar{z}= \sum_{i=1}^{x} z_i / x, 
\end{align}
and $z_i$ denotes the monetary value of each transaction with $x$ as the number of transactions of a customer. Note that this model does not consider time as a dimension. The underlying "true" average transaction value is assumed to be time-invariant.

To be able to model $\bar{z}$, the Gamma-Gamma model assumes that a customer's spending per transaction is Gamma distributed with $p$ as shape and $\nu$ as scale parameter:
\begin{align}
    g(z_i)=\frac{\nu^p z_i^{r-1} e^{-z_i \nu}}{\Gamma(p)}. 
\end{align}
    
Following the Gamma distribution's scaling property, the density of customers' past average spending values $\bar{z}$ is
\begin{align}
    f(\bar{z} | p,\nu;x) & =\frac{(\nu x)^{px}\bar{z}^{px-1}e^{-\nu x \bar{z}}}{\Gamma(px)}. 
\end{align}
    
To allow for heterogeneity in a customer's average spending value, we assume $\nu$ to be Gamma distributed with shape parameter $q$ and the scale parameter $\gamma$. Then, the marginal distribution for the mean spending rate $\bar{z}$ given $x$ is
\begin{align}
    \label{gammagammaLL}
    f(\bar{z} | p,q,\gamma;x) & = \int_{0}^{\infty}\frac{(\nu x)^{px}\bar{z}^{px-1}e^{-\nu x\bar{z}}}{\Gamma(px)}\frac{\gamma^{q}\nu{q-1}e^{-\gamma\nu}}{\Gamma(q)}d\nu \nonumber \\
                              &                                                                                                                                                   
    = \frac{1}{\bar{z}B(px,q)}\Bigg(\frac{\gamma}{\gamma+x\bar{z}}\Bigg)^q\Bigg(\frac{x\bar{z}}{\gamma+x\bar{z}}\Bigg),
\end{align}
where $B(\cdot,\cdot)$ is the Beta function, see, e.g., \citet{abramowitz1964}. The parameters $p$, $q$, and $\gamma$ of Eq. (\ref{gammagammaLL}) are estimated using maximum likelihood.  

Recently, extensions to include covariates in the GG model have been proposed. For details, see the working papers of \citet{Bachmann2022a, Bachmann2022b} and \citet{Fader2024}. 

\begin{figure}[h!]
    \centering
    \vspace{0.1cm}
    \includegraphics[width=0.95\linewidth]{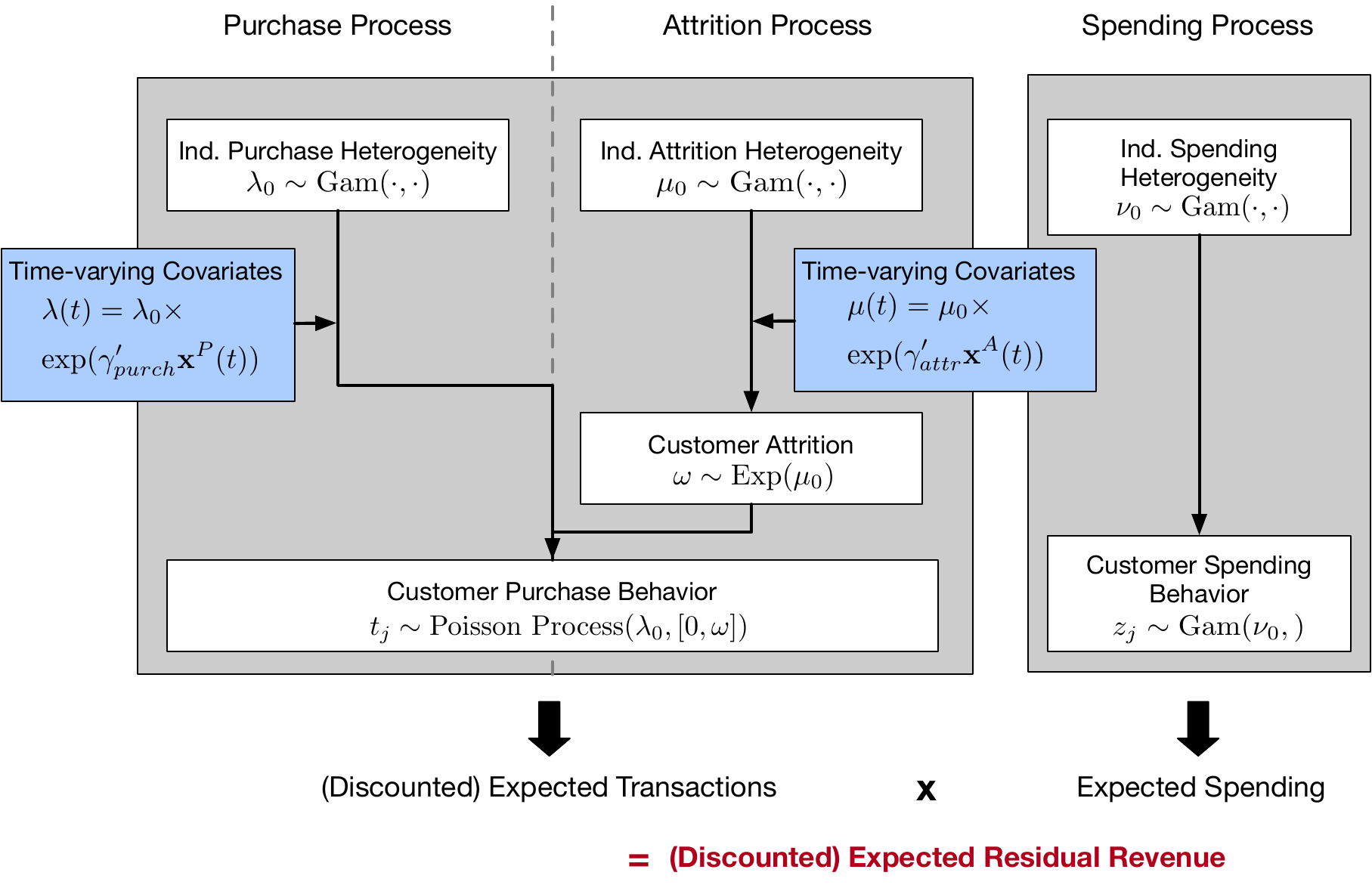}
    \vspace{-0.2cm}
    \caption{General modeling structure -- Pareto/NBD model \& Gamma-Gamma model (Gam refers to the Gamma distribution, and Exp refers to the Exponential distribution.)}
    \label{fig:ePNBDillustration}
\end{figure}

To predict an individual CLV, we multiply a customer's average spending per transaction, as predicted by using the Gamma-Gamma model, with the discounted transactions for this particular customer, as predicted by a latent attrition model. Figure \ref{fig:ePNBDillustration} provides an illustration of the Gamma-Gamma model and its combination with the Pareto/NBD model (with time-invariant and -varying variables) to infer customers' value.

\section{Practical modeling guidance from the literature} \label{guidance}
    
Practical guidelines on probabilistic modeling focus primarily on latent attrition models. These guidelines are based on either theoretical arguments or simulation studies. In particular, extensive guidance is available for the most popular latent attrition model, the Pareto/NBD model. These recommendations mainly refer to the definition of the estimation period, such as its coverage, length, and sample size. Another focus of related studies has been the length of the prediction period. In the following, we provide an overview of these guidelines and point to often overlooked details discussed in previous research \citep[e.g.,][]{Simon2022, Schwartz2014, Hoppe2010}.
    
In general, it is common practice to split the customer base into multiple customer cohorts that are defined based on their first purchase and to fit separate models for each. Some authors refer to these as acquisition cohorts. The theoretical argument underlying this procedure is that these customers are prone to share some characteristics and thus their purchase behavior is often relatively homogeneous. For example, customers acquired in the last quarter of the year, e.g., in the context of Black Friday promotions, may have worse retention than customers acquired in other quarters. Generally speaking, customers attracted to a business may change over time as the market evolves, consumer preferences shift, or the product's brand positioning changes. Acquisition cohorts provide a convenient way to account for such changes.
It can be worthwhile to divide a cohort defined alone by its acquisition time further using other attributes of the first purchase. Common choices are the campaign or channel that are considered to have acquired a customer because they are expected to reach different audiences.
This implicitly assumes that a firm's targeting strategy works, i.e., that a firm's marketing effort indeed resonates with a particular, relatively homogeneous consumer group. The work of \citet{Jaworski1985} provides an in-depth discussion on cohort analysis, including its history in marketing research. However, it is also possible to estimate the entire customer base with a single model controlling for customer cohort assignment by binary indicator variables. These "cohort dummies" control for the cohort-specific effects on the mean purchase rate $\lambda$ and the mean attrition rate $\mu$. Alternatively, practitioners might even prefer a simpler solution and trade-off some predictive accuracy and insights in cohort-specific patterns. In this case, they would estimate a single model for the entire customer base without controlling for customer-cohort-specific effects. An intermediate approach between the use of "cohort dummies" and cohort-specific models is to specify a parametric cross-cohort drift such as including a linear covariate for cohort assignment, which assumes that there is a linear relationship between cohorts of different ages.
    
When customer cohorts are formed, recommendations on minimum cohort size range from 750 to 1,600 \citep{Simon2022, Schmittlein1994}. However, other research also finds that sometimes even "small samples of 250 to 500 customers may provide reliable and sufficiently precise estimates" \citep[p.49]{Hoppe2010}. The ability to estimate the customer base in a single model and use covariates to control for cohorts-specific effects might provide a solution for firms with a smaller customer base, such as in the business-to-business industry. In other scenarios, such an approach also enables analysts to control for more fine-grained cohort effects (e.g., by using weekly cohort dummies) within coarser cohorts (e.g., within a yearly cohort sample).
    
Complementing these insights, recent work recommends analyzing recent cohorts separately and merging older cohorts in a single sample: "Even if you have complete records for every single customer from the day your firm started, you may choose to have a 'pre-20xy' cohort to make the various plots that follow more legible. Besides, long after the shakeout [...] takes place, the distinctions between cohorts are less meaningful and interesting. The benefits arising from tracking old cohorts as separate entities will often not be worth the effort to do so" \citep[p.193]{Fader2022}.
    
Another question is how latent attrition models should \textit{account for the first purchase or not}. Often, this observation is used to identify the assignment to a customer cohort. However, this may not be the best approach for any analysis. For some business contexts, an alternative approach can be better suited. While rarely applied in business practice, this has already been highlighted in the seminal work by \citet{Schmittlein1987}. For example, if data on the sign-up date for a loyalty program is available, it can be used to define the cohort assignment, and the first purchase will be counted in the Pareto/NBD model. This also applies if other customer interactions with the firm precede the first transaction, e.g., a free trial period. Note that this can be different for the spending model (see Section \ref{models}).
    
Analogously, the \textit{length of the estimation period} depends on the business context. If seasonal patterns are not present, the estimation period can be determined based on customers' interpurchase time and dropout rate \citep{Hoppe2010, Simon2022}. In general, we recommend choosing an estimation period that is at the very least as long as the average interpurchase time. Thus, a startup business that only has very few months of purchase records but very short interpurchase times can use these models. For many retailers, it often makes sense to at least look at the last 12 months, as purchasing behavior often fluctuates greatly over the course of the months \citep{Schmittlein1994}. However, the duration of the estimation period should be extended to 24 months when strong seasonal patterns are observed. Thus, these effects can be better identified by the model. However, even if all past transaction records are available, \citet{Schmittlein1987} recommend testing if the use of only recent transactional data leads to better prediction results. Their argument is that Bayesian updating on customers' death rate can drive the attrition rate $\mu$ for some customers close to zero, which can lead to unrealistically long lifetime predictions. 
 
Further, the business context also determines the choice of the relevant \textit{unit of time} (e.g., hour, day, week, month). To define the input data for a latent attrition model, the chosen time unit is used to convert relevant time spans into meaningfully scaled numbers. In other words, the definition of a time unit forms the basis to convert the raw length of measured time spans to a sensible scale. The scaled numbers then serve as model inputs. The choice of time unit is arbitrary, but it can be a convenient tool to facilitate interpretation and improve numerical stability. In practice, a time unit may be chosen such that the typical interpurchase time is a few single-digit time units long. In most settings, this means choosing \code{"weeks"} as time unit.

Defining the \textit{length of the prediction period} is usually a strategic rather than a methodological decision. However, observing common approaches in business practice, it seems rather paradoxical that marketers often use short-term prediction horizons to evaluate the success of a relationship marketing strategy \citep{Kumar2018}. A key aspect of any relationship marketing is the long-term orientation of a firm and management metrics should be aligned with this \citep{Gummesson1987}. For example, the CLV metric is conceptualized as a manifestation of this long-term orientation \citep{Gupta2006}. From a financial perspective, the net present value of the revenue from future transactions can be substantial. For example, assuming a yearly discount rate of 5\% (10\%), the present value of 1 USD in year 5 is 0.78 USD (0.62 USD). The reasons for the mismatch between strategy and tactics when analyzing and managing customer relationships are well known. Marketing scholar Raghuram Iyengar explains that "customer analytics creates value because it helps organizations maximize the lifetime value of the customer (which is longer term), but most business leaders are compensated for optimizing short-term customer profits across products, divisions, or business units" \citep{Stephen2019}. However, this discrepancy is merely a misconception and not necessarily a contradiction. The CLV metric explicitly values short-term transactions more because of its discounting factor, but at the same time also considers a customer's long-term contribution to the business \citep{Blattberg2009}. The same applies to  managerial metrics to assess a customer's future purchase level, such as the discounted expected residual transactions (DERT) metric, that can be derived from probabilistic latent attrition models \citep[e.g.,][]{Fader2005}.

\textit{Covariates} are usually included for both the lifetime and the transaction process. However, if an analyst has access to a multitude of covariates and prediction is the main concern, it could be an option to include key covariates only for the transaction process. Not only does this reduce the number of parameters that have to be estimated, but there can also be substantive reasons for such an approach. For many businesses, the impact of, e.g., seasonality patterns on the transaction process is stronger and more straightforward to determine. Industry experts have taken advantage of the opportunity to include time-dependent covariates to create indicator variables based on the residuals between predictions and actual values. This must be done with the utmost care to avoid overfitting and information leakage, but it can help to gain additional insights, e.g., about the relevant seasonal patterns for a company.  

Modeling \textit{Customers' spending} may also require attention. To account for inflation, users can rely on price indices. Thereby, past monetary values are adjusted to reflect current price levels. To account for costs, it is recommended to rely on an additional model. 

Readers interested in the computational complexity of these probabilistic  models and their various extensions are pointed to Appendix \ref{app:run_times}. Using consumer-grade hardware, we provide some exemplary information on run times for the Pareto/NBD model for 1,000 to 1,000,000 customers. Also, the additional computational complexity of more advanced modeling techniques such as accounting for covariates or correlated processes is discussed. However, it is to note that these results are only valid for the current implementation of the Pareto/NBD model in \pkg{CLVTools} and the hardware used for these analyses. A wide variety of uses case should be able to work with the scalability of the model implementations in \pkg{CLVTools}. For example, we are aware of businesses with 10,000,000 customers that use this implementation of the Pareto/NBD model to estimate and predict the individual future purchase behavior of their customer base.

\section[Using the CLVTools package]{Using the \pkg{CLVTools} package}\label{walkthrough}

To illustrate a common application, we use \pkg{CLVTools} to analyze consumer purchase records from a retail business. The data structure is identical for any type of non-contractual business's transactional records. Thus, the following analyses can be applied "as is" to a broad range of businesses. See Table \ref{table:FunctionalityOverview} for an overview of the key functionalities of \pkg{CLVTools}. 

Our analysis is structured in five sections. First, we discuss data preparation. Thereby, we detail how to initialize and inspect a transaction data object. Second, we discuss model estimation and diagnostics for both latent attrition models and spending models. Third, we explain how to derive predictions of individual customer lifetime values using the models obtained in the previous step. We look at two scenarios: First, we illustrate how to split the transaction data into an estimation and holdout set to fit the model parameters on the former and evaluate performance on the latter. Second, we also illustrate how to make predictions once the analyst has decided on a final model, using all available data. In a fourth section, we illustrate how to account for various types of covariate information. In other words, how to estimate latent attrition models with time-invariant and time-varying covariates and compare their performance to the same model without covariates. Lastly, we show how to leverage advanced modeling technique to deepen the understanding of drivers of purchase behavior. 

\begin{table}[h!]
        
    \renewcommand*{\arraystretch}{1.058}

    \begin{tabular}{@{}p{2.1cm}p{3.75cm}p{4.35cm}p{4.1cm}@{}}
        \toprule
        \textbf{Step}                                                                      & \textbf{Details}                                              & \textbf{Commands}                                    & \textbf{Key arguments}                     \\
        \midrule
        \\[-2ex]
                
        \textit{(1) Data} \newline \textit{preparation}                                    & Create data object \newline from transaction \newline history & \code{clvdata()}                                     & \code{data.transactions, \newline          
        time.unit, \newline
        estimation.split \newline
        data.end
        }  \\
        \\[-1.5ex]
        \cdashline{2-4}
        \\[-1.75ex]
                                                                                           & \textit{For \code{clvdata} objects}                           &                                                      &                                            \\
        \\[-2.25ex]
        \cdashline{2-4}
        \\[-1.5ex]
                                                                                           & Add time-invariant \newline covariate data                    & \code{SetStaticCovariates()}                         & \code{clv.data, \newline                   
        data.cov.life,  \newline
        data.cov.trans
        }   \\[6pt]
                                                                                           & Add time-varying \newline covariate data                      & \code{SetDynamicCovariates()}                        & \code{clv.data, \newline                   
        data.cov.life,  \newline
        data.cov.trans
        } \\[6pt] 
                                                                                           & Data subsetting                                               & \code{subset()}                                      & \code{ids}                                 \\[6pt]	
                                                                                           & Data extraction                                               & \code{as.data.frame(), \newline as.data.table()}     & \code{clv.data \newline sample}            \\[6pt] 
                                                                                           & Data summary                                                  & \code{summary()}                                     & \code{ids}                                 \\[6pt]
                                                                                           & Descriptive plots                                             & \code{plot()}                                        & \code{                                     
        which, 
        sample
        }\\[6pt]  
        \hdashline
        \\[-1.5ex]
        \multirow{2}{2.1cm}{\textit{(2) Model \newline estimation}}                        & Latent attrition model                                        & \code{latentAttrition()}                             & \code{formula\textsuperscript{\dag}, data} \\[6pt]
                                                                                           & Spending model                                                & \code{spending()}                                    & \code{formula\textsuperscript{\dag}, data  
        }\\[6pt]  
        \hdashline
        \\[-1.5ex]
        \multirow{3}{2.1cm}{\textit{(3) Model \newline evaluation \& \newline prediction}} & Prediction routine                                            & \code{predict()}                                     & \code{clv.fitted, newdata}                 \\[6pt]
                                                                                           & Diagnostic plots                                              & \code{plot()}                                        & \code{                                     
        which 
        }  \\[6pt]
                                                                                           & Model summary                                                 & \code{summary()}                                     & \code{clv.fitted}                          \\[6pt]
                                                                                           & Model details                                                 & \code{vcov(), confint(), logLik(), nobs(), fitted()} & \code{clv.fitted}                          \\
        \bottomrule
        \multicolumn{4}{p{.95\textwidth}}{\scriptsize Notes: \textsuperscript{\dag}  For details on how to specify the model formulas, see Section \ref{models}. } 
    \end{tabular}
        
    \caption{Overview of the key functionalities of \pkg{CLVTools}} \label{table:FunctionalityOverview}
\end{table}

\subsection{Data preparation}

\subsubsection{Preparing the transaction data}

\pkg{CLVTools} requires customers' purchase history as input data, provided as \code{data.frame} or \code{data.table}. Every transaction record consists of a purchase date and a customer identifier. Optionally, the value of the transaction may be included. Doing so enables the prediction of individual lifetime values. Without this information, the predictions are limited to the future purchase level of customers, i.e., how many times customers will purchase in the future.  
     
For the following case study, we use transactional records that are representative of purchasing data from a retailer in the apparel industry. For each customer there are three variables: Customer identifier (\code{Id}), date of purchase (\code{Date}), and the total monetary value of the transaction (\code{Price}). Data of such structure should be available for any business operating in a non-contractual setting.  We use the command \code{data("apparelTrans")} to load this data as a \code{data.table} object into the working environment.
     
\begin{CodeChunk}
\begin{CodeInput}
R> library(CLVTools)
R> library(data.table)
R> library(ggplot2)
R> library(egg)
R> data("apparelTrans")
R> head(apparelTrans, n = 3)
\end{CodeInput}
\begin{CodeOutput}
        Id         Date    Price
    <char>       <Date>    <num>
1:       1   2005-01-02   230.30
2:       1   2005-09-06    84.39
3:       1   2006-01-18   131.07
\end{CodeOutput}
\end{CodeChunk}

Probabilistic models for customer base analysis, such as those implemented in \pkg{CLVTools}, are often applied per customer acquisition cohort. Such a procedure accounts for possible differences between customers acquired at different time points. The \code{apparelTrans} data represents a single such cohort. It consists of 600 customers who purchased for the first time from this business on the day of 2005-01-02. The modeling can be divided in two stages: 
\begin{itemize}
\item First, during the model benchmarking stage, the purchase records of all customers are split at a set date into two parts, an estimation and a holdout period. The model is estimated solely on the data in the estimation period while the data from the holdout period allows the model's out-of-sample performance to be assessed. The optimal duration of the estimation period depends on the characteristics of the analyzed dataset, while the length of the prediction period is often a strategic management question. Generally, the estimation periods should be at least as long as the average time between customers' transactions in order to observe repeat purchasing. If significant seasonality patterns are present, it is recommended to use at least a period of two years for model estimation (see Section \ref{guidance}). 
\item Second, after assessing model performance with an estimation and a holdout set, one can proceed to estimate a final model on available data and use it to derive the final predictions. For this last step, the model specification that has proven to be most robust in terms of accuracy and reliability during the benchmarking phase is used.

\end{itemize}

For model benchmarking, we first create a transaction data object in which the purchase history is separated into an estimation and holdout set. The resulting object contains both the estimation and holdout data. We use \code{estimation.split} to set the estimation period to 104 periods. This means, 104 weeks since the first purchase record. Alternatively, a date can be provided. By default, the last purchase record marks the end of the data. Optionally, we could use the parameter \code{data.end} to specify a fictional end beyond this date. This is useful, for example, when the last purchase record was on 2000-12-29 but customers were actually observed until 2000-12-31. Using \code{data.end="2000-12-31"} without holdout period defines the estimation period until 2000-12-31 and moves the start of the prediction period to 2001-01-01.
\begin{CodeChunk}	
\begin{CodeInput}
R> clv.apparel <- clvdata(
+     data.transactions = apparelTrans,  
+     date.format = "ymd", 
+     time.unit = "week",
+     estimation.split = 104,
+     name.id = "Id", 
+     name.date = "Date",
+     name.price = "Price"
+ )
\end{CodeInput}
\end{CodeChunk}

Further, we define the argument \code{time.unit}, here by specifying the value \code{"week"}. This period length is used to measure time spans. Further, we define the argument \code{date.format}, here by choosing \code{"ymd"} for year-month-day. This format is used to parse the date column in the given data, but also for any other date-like input at a later stage, such as parameters given as characters. The time-unit definition also prescribes how the transaction log must be aggregated before estimation. When the user creates a transaction data object in \texttt{CLVTools}, two aggregation regimes apply. If unit of time for the probabilistic model is defined as \texttt{"daily"}, \texttt{"weekly"}, or \texttt{"yearly"}, the package first collapses the data to the day level: For any customer–day combination, multiple purchases are combined into a single record whose transaction count equals one and whose monetary value equals the sum of that day’s spending. For all other settings of \texttt{time.unit}, e.g., hours, \texttt{CLVTools} aggregates at the most granular resolution allowed by a \texttt{POSIXct} timestamp - one second - so that at most one observation per customer–second remains. Any coincident purchases are merged -- that is, the respective count is set to one -- and their monetary values are summed. This automatic preprocessing has the following reason: Often purchases on the same day should not be regarded as separate, independent events but as strongly linked, e.g., a customer forgot to purchase milk and goes back into the grocery store after paying for all other errands. The Poisson process explicitly assumes that the events are independent. Because of this aggregation, purchases will not be stored at a finer temporal resolution than seconds for \code{time.unit=hours} and days for all other time units. Consequently, the time between purchases can only be measured at this granularity, e.g., at steps of 1/7 for \code{time.unit = "week"}. This preprocessed data input then serves as input to both latent attrition models and spending models.

As an alternative to \code{clvdata()}, the command \code{as.clv.data()} is available to convert a \code{data.frame} directly. It uses the most common defaults for all arguments.

\subsubsection{Inspecting the transaction data}

Customers' purchase records can be extracted using the generics \code{as.data.frame()} and \code{subset()}. 
By default, \code{as.data.frame()} copies all data present in the object, but this can be limited.  With the parameter \code{sample}, data can be selected from either the estimation set (\code{"estimation"}), the holdout set (\code{"holdout"}), or both (\code{"full"}). The parameter \code{ids} restricts the purchase records to the customers with matching identifiers. 

\begin{CodeChunk}
\begin{CodeInput}
R> df.full <- as.data.frame(clv.apparel, sample = "full") 
R> df.est <- as.data.frame(clv.apparel, sample = "estimation") 
R> df.ids <- as.data.frame(clv.apparel, ids = "1") 
\end{CodeInput}
\end{CodeChunk}	

For more detailed data selection, the \code{subset()} method is available. It returns the transaction data that meets the specified conditions. The given conditional expressions are applied on the internal \code{data.table} object that holds the purchase data. The usual \code{data.table} syntax can hence be used to select based on the columns in the internal data: \code{Id}, \code{Date}, and, if provided, \code{Price}. Additionally, the argument \code{sample} is available again.

\begin{CodeChunk}
\begin{CodeInput}
R> df.price.50.to.100 <- subset(clv.apparel, Price >= 50 & Price <= 100)
R> df.ids.7.9 <- subset(clv.apparel, Id=="7"|Id=="9", sample = "holdout")
\end{CodeInput}
\end{CodeChunk}

The \code{summary()} method provides descriptive statistics for the transaction data. 
     
\begin{CodeChunk}
\begin{CodeInput}
R> summary(clv.apparel)
\end{CodeInput}
\begin{CodeOutput}
CLV Transaction Data 
                                
Time unit         Weeks         
Estimation length 104.0000 Weeks
Holdout length    207.0000 Weeks

Transaction Data Summary 
                                  Estimation     Holdout        Total
Period Start                      2005-01-02     2007-01-01     2005-01-02
Period End                        2006-12-31     2010-12-20     2010-12-20
Number of customers               -              -              600       
First Transaction in period       2005-01-02     2007-01-01     2005-01-02
Last Transaction in period        2006-12-31     2010-12-20     2010-12-20
Total # Transactions              1866           1317           3183      
Mean # Transactions per cust      3.110          5.557          5.305     
(SD)                              2.714          5.123          6.119     
Mean Spending per Transaction     40.545         36.977         39.069    
(SD)                              73.362         55.356         66.519    
Total Spending                    75657.730      48699.170      124356.900
Total # zero repeaters            213            -              -         
Percentage of zero repeaters      35.500         -              -         
Mean Interpurchase time           24.823         30.604         37.817    
(SD)                              19.417         24.756         42.339  
\end{CodeOutput}

\end{CodeChunk}

For this dataset, the summary statistics show, for example, that the dataset includes a total of 600 customers who made 3'183 purchases. 35.5\% of these customers are zero repeaters, i.e., they purchase only once and never return. 

Through the \code{plot()} method, a variety of descriptive plots are available. These can be selected using the argument \code{which}. Table \ref{table:ClvdataPlots} provides an overview of the available plots. 

\begin{table}[h!]
    \centering
    \vspace{0.5cm}
    \begin{tabular}{p{4.2cm}p{4.2cm}p{6cm}}
        \toprule
        \\[-2.25ex]
        Plot                               & \code{which}                 & Description                                                         \\
        \\[-2ex]
        \midrule
        \\[-1.5ex]
        {\textit{Tracking Plot}}           & {\code{"tracking"}}          & The total number of repeat transactions per period.                 \\
        \\[-1.5ex]
        {\textit{Frequency Plot}}          & {\code{"frequency"}}         & The distribution of the number of transactions per customer.        \\
        \\[-1.5ex]
        {\textit{Spending Plot}}           & {\code{"spending"}}          & The empirical density of the average spending per transaction.      \\
        \\[-1.5ex]
        {\textit{Interpurchase Time Plot}} & {\code{"interpurchasetime"}} & The empirical density of customer's mean time between transactions. \\
        \\[-1.5ex]
        {\textit{Transaction Timing Plot}} & {\code{"timings"}}           & The transaction timings of a customer subset.                       \\
        \\[-1.5ex]
        \bottomrule
    \end{tabular}
    \caption{Descriptive plots available for transaction data objects}
    \label{table:ClvdataPlots}
        
\end{table}

\begin{figure}[h!]
    \centering
    \includegraphics[width=1\textwidth]{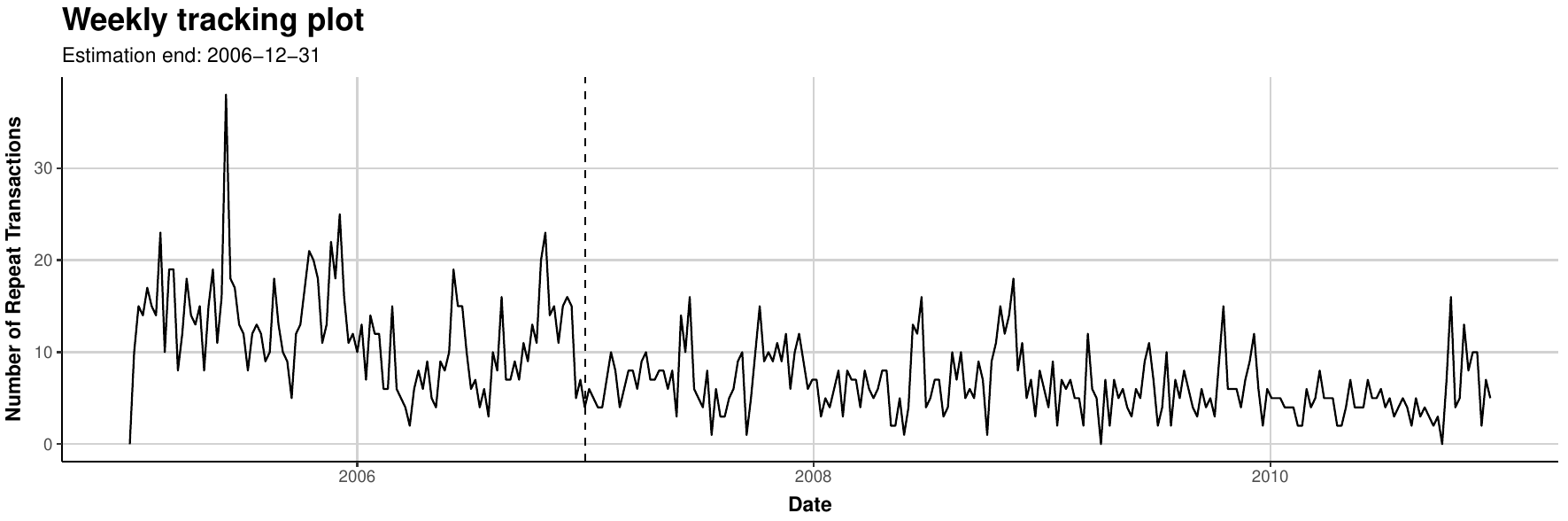}
    \caption{Tracking plot for the apparel dataset}
    \label{plot:tracking}
        
\end{figure}

The tracking plot (\code{which = "tracking"})  -- see Figure \ref{plot:tracking} -- shows the total number of repeat transactions observed in each period. It helps to spot potential seasonal patterns, and thus informs the analyst whether model extensions to control for these patterns should be explored. 

\begin{CodeChunk}
\begin{CodeInput}
R> plot(clv.apparel, which = "tracking")
\end{CodeInput}
\end{CodeChunk}

The tracking plot illustrates the number of repeat purchases, i.e., excludes the first purchase of every customer. Each point in the plot represents the number of repeat orders that were placed in the time period preceding this date. The data's very first order defines the start of the plot. Because there is no time to make any repeat purchases until then, this initial value by definition is zero and gives this plot its characteristic shape for the first time period. 
    
Next, the timings plot (\code{which="timings"}) helps to visualize customers' transaction timings. 

\begin{CodeChunk}
\begin{CodeInput}
R> plot(
+     clv.apparel, 
+     which = "timings", 
+     annotate.ids = TRUE,
+     ids = c("1", "2", "3")
+ )
\end{CodeInput}
\end{CodeChunk}

\begin{figure}[h!]
    \centering
    \includegraphics[width=1\textwidth]{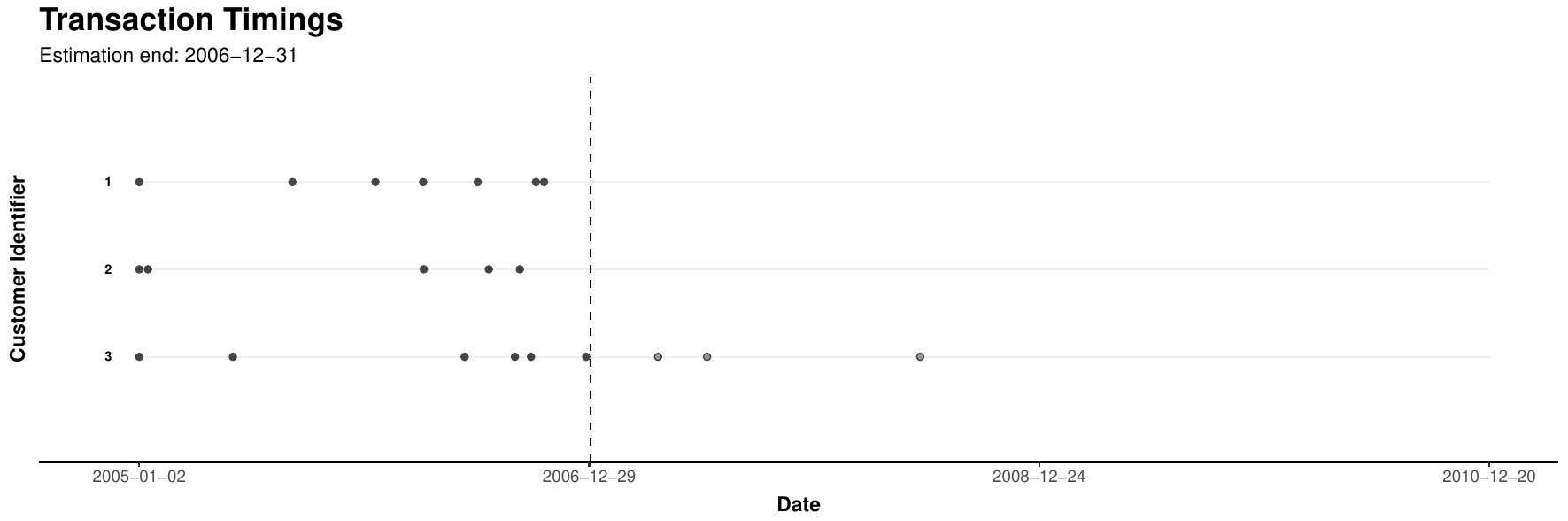}
    \caption{Timings plot for the apparel dataset}
    \label{plot:timings}
     
\end{figure}

The parameter \code{ids} can be either a vector of strings listing a subset of customers or an integer specifying the number of customers to randomly select. The resulting plot is shown in Figure \ref{plot:timings}. Each dot in a row illustrates when a transaction for a particular customer was recorded. It provides an intuitive way to learn about the purchase patterns of individual customers. 

The interpurchase-time plot (\code{which="interpurchasetime"}) provides additional insights. 

\begin{CodeChunk}
\begin{CodeInput}
R> plot(clv.apparel, which = "interpurchasetime", sample = "estimation", bw=3)
\end{CodeInput}
\end{CodeChunk}

\begin{figure}[h!]
    \centering
    \includegraphics[width=1\textwidth]{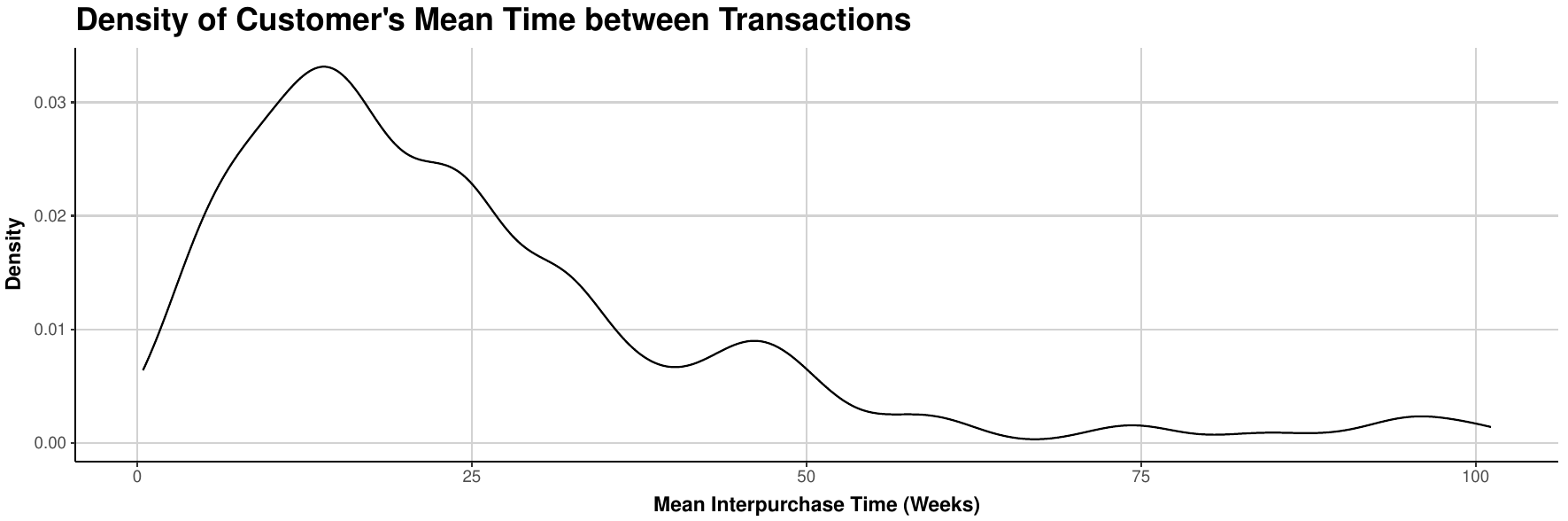}
    \caption{Interpurchase-time plot for the apparel dataset}
    \label{plot:interpurchasetime}
\end{figure}

The resulting plot is shown in Figure \ref{plot:interpurchasetime}.
It illustrates the empirical density of customers' mean time between transactions, after aggregating purchases of the same customer on the same date. The time unit defined in the data object is used to measure the time between transactions. Only data from customers with repeat transactions are shown in this graph. This plot can help to visually inspect the distribution of interpurchase times and thus verify model assumptions. Note that the plot interface allows us to pass additional arguments that are forwarded to \code{ggplot2::stat_density}, e.g., to adjust the bandwidth (\code{bw=3}).

If price information is present in the data object, the spending plot (`\code{which = "spending"}) is also available. It either shows the empirical density of customers' average order values (\code{mean.spending = TRUE}) or the value of every transaction in the data (\code{mean.spending = FALSE}). Such a plot can notably be used to assess whether the distribution remains stable in the estimation and the holdout period, a key assumption of the spending models.

\begin{CodeChunk}
\begin{CodeInput}
R> plot.7a <- plot(clv.apparel, which = "spending", sample = "estimation") + 
+             xlim(0,550)
R> plot.7b <- plot(clv.apparel, which = "spending", sample = "holdout") + 
+             xlim(0,550)
R> ggarrange(plot.7a, plot.7b, labels = c("A", "B"), ncol = 2, nrow = 1)
\end{CodeInput}
\end{CodeChunk}

\begin{figure}[h!]
    \centering
    \includegraphics[width=1\textwidth]{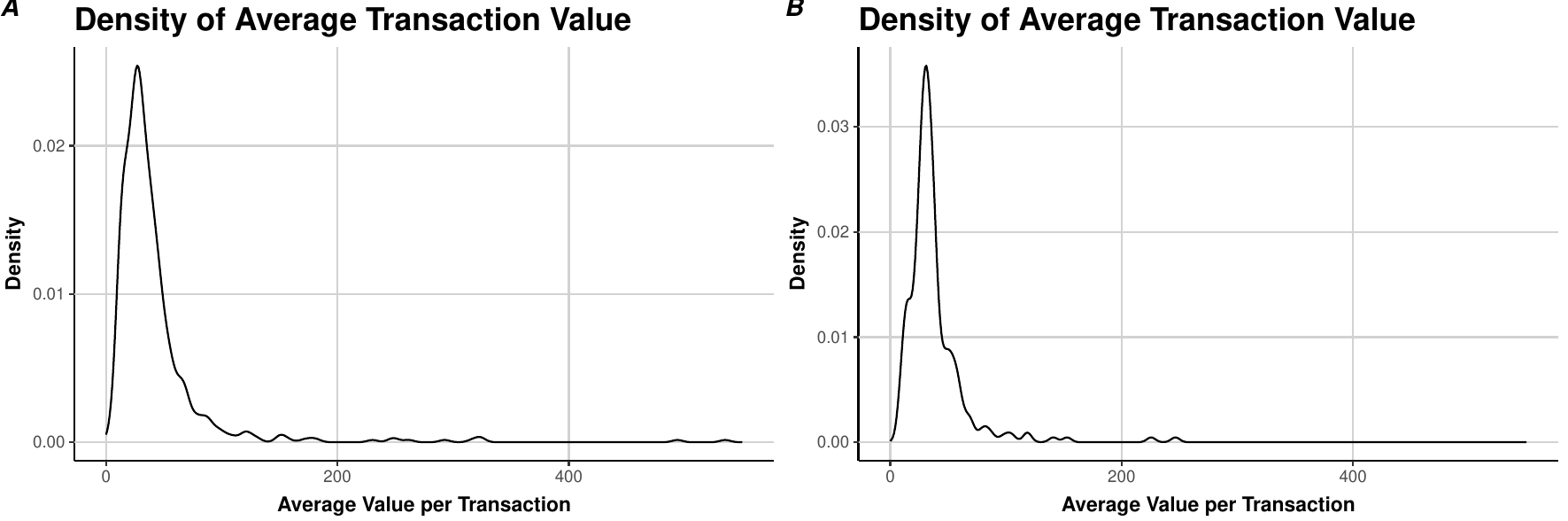}
    \caption{Spending plots for the apparel dataset (A: estimation period; B: holdout period)}
    \label{plot:spending1}
\end{figure}

The resulting plots are shown in Figure \ref{plot:spending1}. As for the interpurchase-time plot, the density bandwidth (\code{bw}) can be passed as an additional argument. For the data of the case study, the densities of customer spending in the estimation and holdout period are quite similar. We note that higher average transaction values are observed at the extreme end of the estimation period. This provides support for the probabilistic approach's assumption that the spending patterns of customers remain constant over time.

\subsection{Model estimation and diagnostics}\label{models}

\subsubsection{Fitting a latent attrition model}

After creating the transaction data object, we estimate a first latent attrition model on it. We focus on the Pareto/NBD model \citep{Schmittlein1987}, which has proven to be very robust across various industries. As with any customer base analysis, our case study starts with the simplest case. This means that we do not include any covariates in our first model. Recall that the required model inputs $(x_i, t_{x_i}, T_i)$ can be derived from the purchase history of the estimation period alone, and no other data about the customers or holdout period is needed. 

We use the \code{latentAttrition()} interface to estimate the model. Besides the purchasing data (\code{data}), we specify the model family (\code{family = pnbd}). The parameter \code{formula} is used to specify the covariates. Because we are estimating a model without covariates, this parameter has to be omitted. Recall that for models without covariates, all inputs are predefined and therefore, no dependent variable of any sort needs to be specified either.

\begin{CodeChunk}
\begin{CodeInput}
R> est.pnbd <- latentAttrition(family = pnbd, data = clv.apparel)
R> est.pnbd
\end{CodeInput}
\begin{CodeOutput}
Pareto/NBD Standard Model

Call:
latentAttrition(family = pnbd, data = clv.apparel)

Coefficients:
r     alpha        s      beta  
1.4490   48.6361   0.5613   46.8844  
KKT1: TRUE 
KKT2: TRUE 

Used Options:
Correlation:     FALSE
\end{CodeOutput}
\end{CodeChunk}

The number and names of the reported parameters vary by model. For the Pareto/NBD model, we obtain four parameters $r, \alpha, s$ and $\beta$. $r$ and $\alpha$ represent the shape and scale parameters of the distribution that determines the purchase rate and $s$ and $\beta$ represent the shape and scale parameters of the distribution that determines the attrition rate across individual customers. $r/\alpha$ can be interpreted as the mean transaction and $s/\beta$ as the mean attrition rate. In the case of the apparel dataset, we observe an average purchase rate of $r/\alpha=0.030$ transactions and an average attrition rate of $s/\beta=0.012$. 

Use \code{summary(est.pnbd)} to obtain a more detailed report that includes the likelihood value as well as AIC and BIC. Further, it includes the Karush-Kuhn-Tucker optimality conditions of the first (KKT1) and second order (KKT2) \citep{KKT}. Care needs to be taken, if the first KKT condition is not true, as this might hint at a possible model misspecification.

If questions on optimality arise, the parameter \code{optimx.args} allows the optimization routine to be controlled. 

\label{code:optimxargs}
\begin{CodeChunk}
\begin{CodeInput}
R> est.pnbd <- latentAttrition(
+     family = pnbd, 
+     data = clv.apparel,
+     optimx.args = list(method = "Nelder-Mead", control = list(trace = 1))
+ )
\end{CodeInput}
\end{CodeChunk}

\noindent The argument \code{optimx.args} forwards a list of arguments to \pkg{optimx} \citep{optimx1, optimx2}. This allows us to define the optimization algorithm, set upper or lower limits, or trace the optimization progress. \pkg{CLVTools} by default uses the optimization method \code{L-BFGS-B} \citep{byrd1995limited}. If the optimization is not feasible, switching to the more robust but often slower method \code{Nelder-Mead} is recommended \citep{nelder1965simplex}. 

As an alternative to the Pareto/NBD model, \pkg{CLVTools} features the Beta-Geometric/NBD model \citep{Fader2005c} and the Gamma-Gompertz/NBD \citep{Bemmaor2012} model. To use these models, set the parameter \code{family} to either \code{bgnbd} or to \code{ggomnbd}. Table \ref{table:ModelsOverview} gives an overview of the available latent attrition models and the implemented advanced modeling options. 

\begin{table}[h!]
    \centering
    \vspace{0.5cm}
        
    \begin{tabular}{p{5.4cm}p{2.8cm}p{2.8cm}p{2.8cm}}
                
        \toprule
                
        \\[-2.25ex]
                
                                                                                    & \code{pnbd}                    & \code{bgnbd}                 & \code{ggomnbd}              \\
                
        \\[-2ex]
                
        \midrule
                
        \\[-1.5ex]
                
        {\textit{Model name}}                                                       & {Pareto/ \newline NBD}         & {BG/ \newline\ NBD}          & {GGom/ \newline NBD}        \\
                
        \\[-1.5ex]
                
        {\textit{Attrition: Distribution \newline main process \& heterogeneity}}   & {Exponential / \newline Gamma} & {Geometric /~ \newline Beta} & {Gompertz / \newline Gamma} \\
                
        \\[-1.5ex]
                
        {\textit{Transaction: Distribution \newline main process \& heterogeneity}} & {Poisson / \newline Gamma}     & {Poisson /~ \newline Gamma}  & {Poisson / \newline Gamma}  \\
                        
        \\[-1.5ex]
        
        \textit{Model Parameters}                                                   & {$r, \alpha, s, \beta$}        & {$r, \alpha, a, b$}          & {$r, \alpha, \beta, b, s$}  
                        
        \\[-1.5ex]
                
        {\textit{Covariates \newline  (time-invariant/-varying)}}                   & \checkmark /  \checkmark       & \checkmark / -               & \checkmark / -              \\
                
        \\[-1.5ex]
                
        {\textit{Process correlation}}                                              & \checkmark                     & -                            & -                           \\
                
        \\[-1.5ex]
                
        {\textit{Equality constraints of \newline covariate parameters}}            & \checkmark                     & \checkmark                   & \checkmark                  \\
                
        \\[-1.5ex]
                
        \textit{Regularization of covariate  \newline parameters}                   & \checkmark                     & \checkmark                   & \checkmark                  \\
                
        \\[-1.5ex]
                
        \bottomrule
                
    \end{tabular}
    
    \caption{Commands and characteristics of latent attrition models in \pkg{CLVTools}}
    \label{table:ModelsOverview}
    
\end{table}

In general, the user can choose between a formula- and a non-formula-based interface to define the model estimation details. Throughout this paper, we illustrate the formula-based interface which is particularly useful for specifying covariates. Alternatively, a model can also be estimated by applying the model family method on the purchasing data object. See the accompanying vignette of the package for more details on this approach.

\subsubsection{Assessing diagnostics for a latent attrition model}

The key diagnostics for a latent attrition model are two plots: (1) the tracking plot and (2) the probability mass function (PMF) plot. Besides these plots and the \code{summary()} command, the canonical generics \code{coef()}, \code{logLik()}, \code{confint()},  \code{vcov()},  \code{nobs()}, \code{AIC()}, \code{BIC()} are available for all fitted models to extract key information.

\begin{figure}[h!]
    \centering
    \includegraphics[width=1\textwidth]{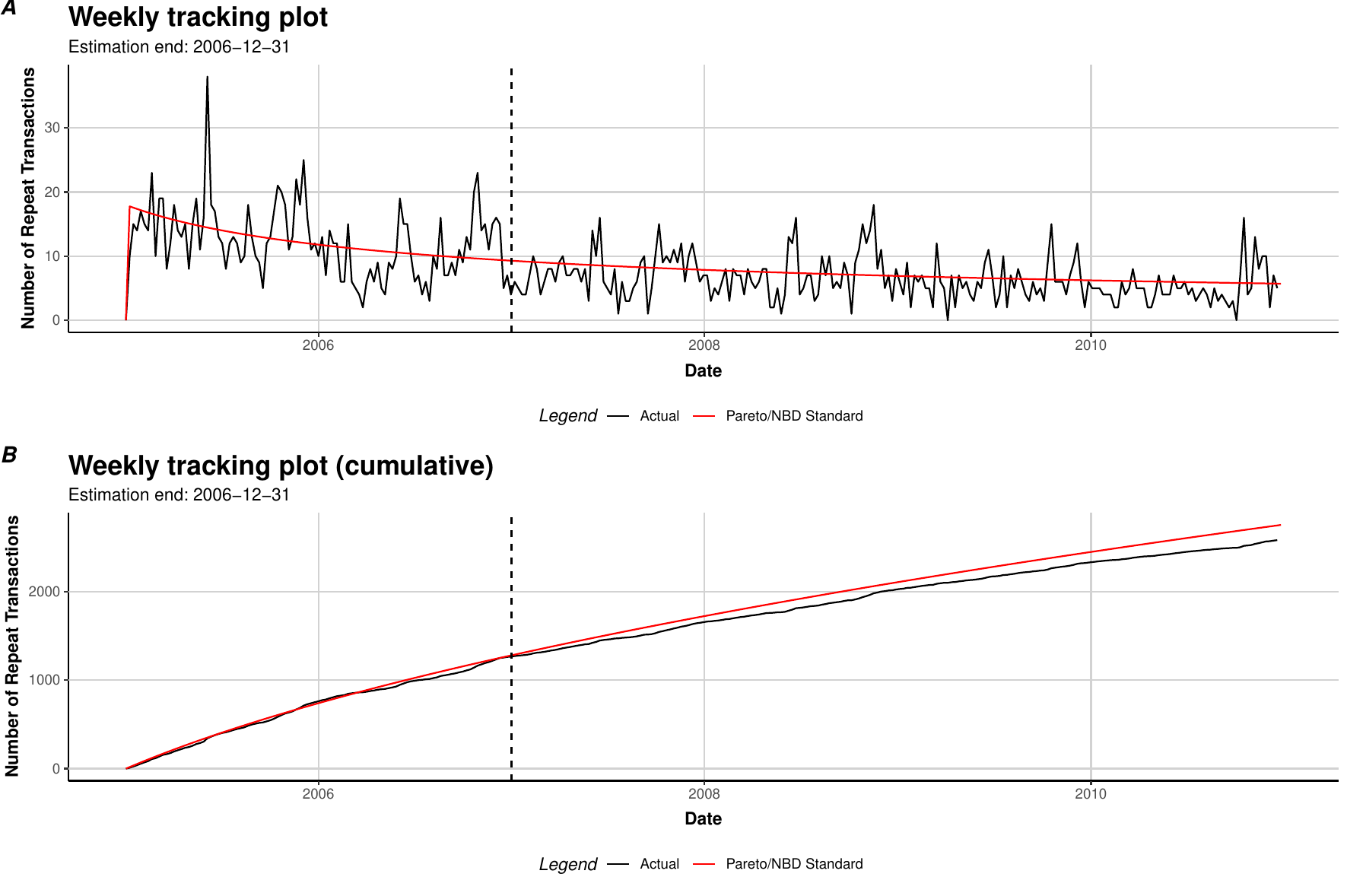}
    \caption{Tracking plots for the apparel dataset (actual versus model results)}
    \label{plot:tracking-model}
\end{figure}

To create a tracking plot, pass \code{which="tracking"} to \code{plot()}. Further, use the flag \code{cumulative} to specify the type of the tracking plot. The plot is drawn until the given \code{prediction.end}. For convenience, the latter defaults to the end of the transaction data. The resulting tracking plots - non-cumulative and cumulative - are shown in Figure \ref{plot:tracking-model}. The dashed line indicates the split between the estimation period and the holdout period. 

\begin{CodeChunk}
\begin{CodeInput}
R> plot.8a <- plot(est.pnbd, which="tracking")
R> plot.8b <- plot(est.pnbd, which="tracking", cumulative = TRUE)
R> ggarrange(plot.8a, plot.8b, labels = c("A", "B"), ncol = 1, nrow = 2)
\end{CodeInput}
\end{CodeChunk}

The tracking plot gives an impression of how well the model tracks the actual purchases in both the estimation and the holdout period. The plot shows the actual repeat transactions and adds an overlay with the repeat transaction expected by the fitted model. This is based on the model's unconditional expectation, i.e., the sum of all transactions expected by all customers in each period \citep[e.g.,][]{Fader2005b}. In line with previous literature, the first predicted date is the start of the data. The expected number of repeat transactions on this date by definition is zero and this fact gives the plot its characteristic shape.

The non-cumulative tracking plot (A) gives an indication of how well the model captures dynamic effects such as seasonality and one-time events. The slightly downward sloping line shows how the model expects fewer purchases over time as more customers stop doing business with the firm.
In the cumulative tracking plot (B) on the other hand, it is easier to see how well the model tracks the number of repeat purchases overall. In this case,  we learn from both plots that the model fits the data quite well. 

\begin{figure}[h!]
    \centering
    \includegraphics[width=1\textwidth]{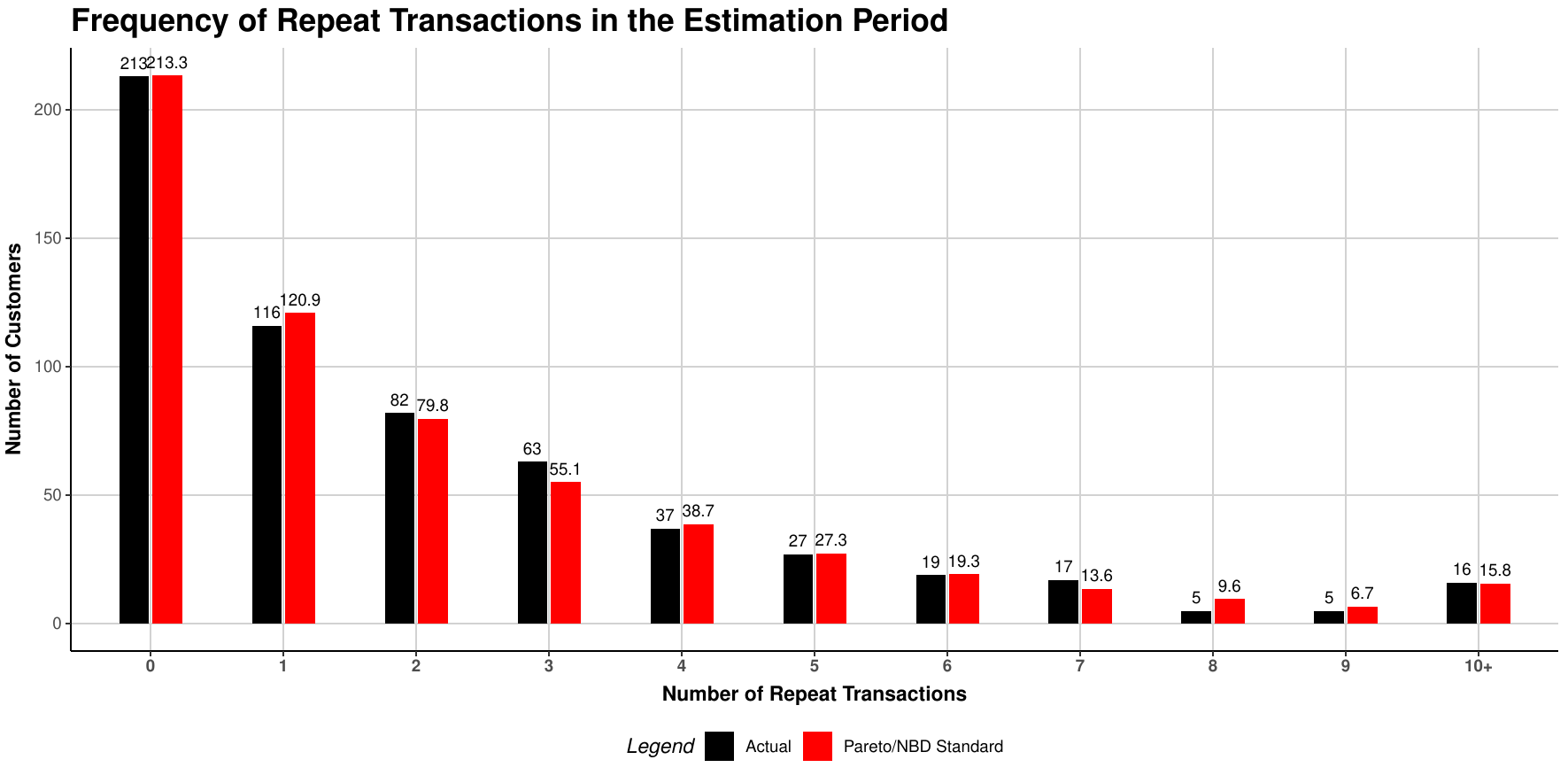}
    \caption{PMF plot for the apparel dataset (actual versus model results)}
    \label{plot:pmf}
\end{figure}

To create a PMF plot, set \code{which="pmf"} in the \code{plot()} command. The resulting plot is shown in Figure \ref{plot:pmf}. The PMF plot completes the assessment of the model's in-sample fit. It shows the actual and expected number of customers who make a given number of repeat transactions during the estimation period. The expected number is based on the PMF, i.e., the probability of making exactly a given number of repeat transactions in the estimation period. For each bin, the expected number of customers is the sum of all customers' individual PMF values for this number of purchases. The comparison between actual and fitted values is straightforward. Again, the results illustrate that the model fits the data well for this application case.

\begin{CodeChunk}
\begin{CodeInput}
R> plot(est.pnbd, which="pmf")
\end{CodeInput}
\end{CodeChunk}

\subsubsection{Fitting a spending model} 

To predict CLV, we need to complement a latent attrition model with a model that estimates customers' average spending per transaction. Currently, only the Gamma-Gamma (GG) model \citep{Fader2005b, Colombo1999} is implemented for which no covariate extension has been published. The Normal-Normal model \citep{Schmittlein1994} is an alternative but the GG model is regarded as more flexible. In literature and practice, it is thus the de-facto standard choice for modeling customers' average spending per transaction.

The specification of spending models is analogous to latent attrition models. There is, however, no parameter \code{formula} as the GG model does not support covariates and the dependent variable is not observed but latent. Should extensions for modeling time-invariant and -varying covariates become available in the future, a formula interface will be added. All required model inputs, in this case, the number of purchases and the mean order value, can again be derived only from customers' purchase history.

In the following, we estimate a GG model using the command \code{spending}. We define the type of model (\code{family}) and the input data (\code{data}). 

\begin{CodeChunk}
\begin{CodeInput}
R> est.gg <- spending(family = gg, data=clv.apparel)
R> est.gg
\end{CodeInput}
\begin{CodeOutput}
Gamma-Gamma Model

Call:
spending(family = gg, data = clv.apparel)

Coefficients:
p       q   gamma  
3.099   5.654  56.504  
KKT1: TRUE 
KKT2: TRUE
\end{CodeOutput}
\end{CodeChunk}

In line with the literature, \pkg{CLVTools} by default does not use the first transaction when estimating a spending model because in many cases this transaction has been found to be atypical for future purchases. As a consequence, customers with a single purchase are ignored during model estimation and their estimated average order value is the mean under the distribution. However, it is possible to set \code{remove.first.transaction=FALSE} to consider the first transaction,.

\subsubsection{Assessing diagnostics for a spending model}

A single diagnostic plot is available for spending models. Additionally, the same canonical generics as for latent attrition models are also available for spending models and help to extract key metrics.

\begin{figure}[h!]
    \centering
    \includegraphics[width=1\textwidth]{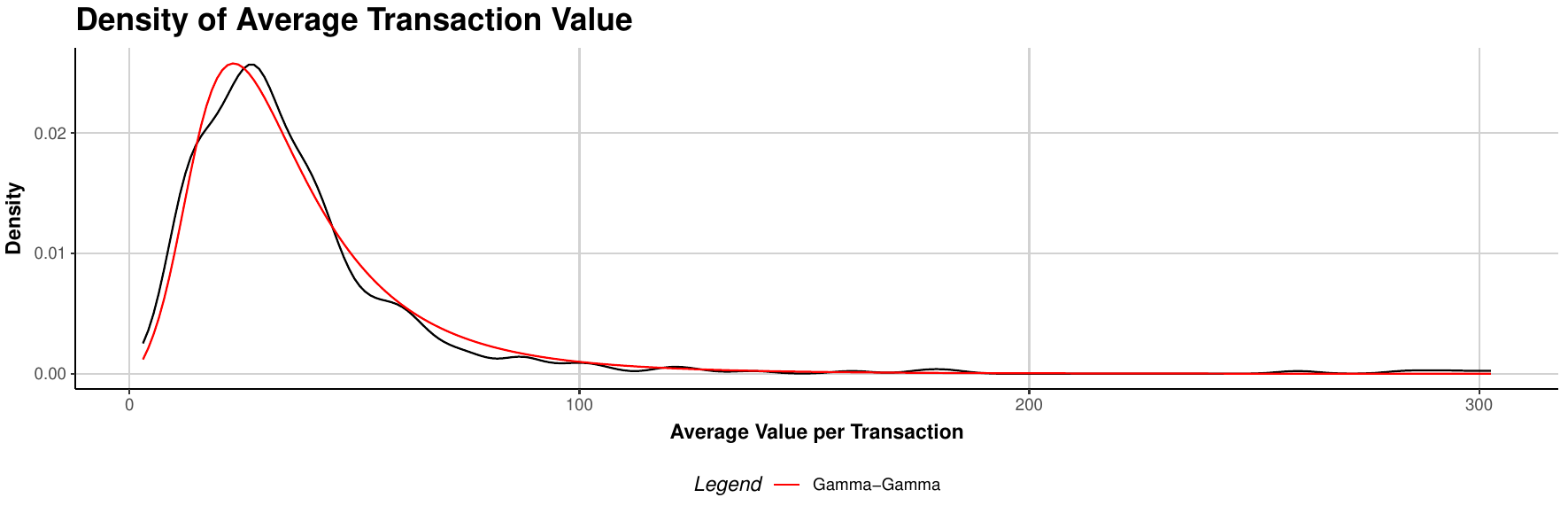}
    \caption{Spending plots for the apparel dataset (actual versus model results)}
    \label{plot:spending2}
\end{figure}

The resulting plot in Figure \ref{plot:spending2} compares the density of each customer's observed average order value (i.e., the empirical distribution) to the model's distribution of mean transaction spending across customers. In this case, the plot shows that the spending model fits the data in the estimation period reasonably well. 

\begin{CodeChunk}
\begin{CodeInput}
R> plot(est.gg)
\end{CodeInput}
\end{CodeChunk}

\subsection{Predicting customer lifetime value}\label{chapter:prediction}

\subsubsection{Evaluating predictions on holdout data}

Once the latent attrition model and the spending model are estimated, we are able to predict various metrics of customers' purchase behavior by using \code{predict()} on the fitted model objects (i.e., \code{est.pnbd} and \code{est.gg}). In order to benchmark different model specifications out-of-sample, we have split the transaction data into an estimation and a holdout period. This allows comparing model predictions for the holdout period to the actuals. We illustrate this in the following but before that we also explain how the predictions of both models are combined.

Recall that latent attrition models predict customers' future number of orders, while spending models infer the average value per order for each customer. Their predictions therefore need to be combined to obtain a measure of future spending and of customer lifetime value. \code{CLVTools} facilitates this by accepting a fitted spending model as an input for the prediction routine of the latent attrition model. The predictions from both models are combined and metrics based on both are calculated. The content of this combined output is outlined in the following. 

Latent attrition models predict three managerial expressions for every customer: \renewcommand{\labelenumi}{(\arabic{enumi})}
\begin{enumerate}
    \item the "conditional expected transactions" (CET), which is the number of transactions to expect from the end of the estimation period until the end of the prediction period,
    \item the "probability of a customer being alive" (PAlive) at the end of the estimation period, and
    \item the "discounted expected residual transactions" (DERT) for every customer, which is the total number of transactions for the residual lifetime of a customer discounted to the end of the estimation period. Note that this metric has an infinite prediction horizon, i.e., the remaining lifetime of each customer is fully taken into account. 
\end{enumerate}

The prediction period can be specified by \code{prediction.end}, either as a duration or as an end-date (i.e, \code{\code{prediction.end = 30} or prediction.end = "2006-05-08"}). If there is a holdout period, the prediction is made by default until the end of the holdout period. The parameter \code{continuous.discount.factor} allows the discount rate used for the DERT expression to be adjusted. 

Spending models alone produce only a single metric for each customer that is independent of a prediction period, namely
\begin{enumerate}
    \setcounter{enumi}{3}
    \item the predicted average spending per transaction.
\end{enumerate}

The combined prediction output of latent attrition and spending models additionally contains
\begin{enumerate}
    \setcounter{enumi}{4}
    \item the customer lifetime value (CLV), which is calculated by multiplying DERT with the average spending per transaction \citep{Colombo1999, Fader2005b}, and 
    \item the total spending expected in the prediction period, which is calculated by multiplying CET with the average spending per transaction. 
\end{enumerate}

Further, if the predictions are made no further than the end of the holdout period, the true number of transactions ("actual.x") and true spending ("actual.total.spending") during the prediction period are reported. This allows for a convenient evaluation with common metrics such as the root mean square error (RMSE) and the mean absolute error (MAE).

Although uncommon in the marketing literature, it is technically also possible to use the DECT predictions, but not the DERT predictions, to assess predictive performance based on the observed data in the holdout period. However, this requires discounting of the observed transactions prior to calculating the RMSE or MAE metric. In addition, this would imply that prediction errors in the far future are weighted less than prediction errors in the near future.

In the following, we now show how to make the combined predictions of both models and evaluate their performance based on the most popular approach in the literature. 
To this end, we first call \code{predict()} on the latent attrition model and additionally pass the fitted spending model by using the argument \code{predict.spending}. Internally, predictions are produced for both models separately and then combined.
Note that both types of model can also be predicted separately. As noted above, the prediction period defaults to the holdout period. 

\begin{CodeChunk}
\begin{CodeInput}
R> dt.pred <- predict(est.pnbd, predict.spending = est.gg)  
\end{CodeInput}
\end{CodeChunk}

If \code{predict.spending=FALSE}, only the predictions for the latent attrition model are produced. For convenience, a spending model can be given (e.g., \code{predict.spending=gg}) in which case the model is estimated and used for prediction. However, we recommended a separate estimation. 

Further, to use the parameters of the estimated models to make predictions for a different set of customers (e.g., for another customer cohort), a new transaction data object can be provided in the argument \code{newdata}. The predictions for these new customers are then made by using the parameters of the estimated model while the inputs required per customer are derived from the transaction log in \code{newdata}.

We now evaluate the models' out-of-sample performance on the holdout data by calculating RMSE and MAE. For such an assessment, the CET metric has to be used. The DERT metric, which is the basis for CLV, incorporates a continuous discount factor and has an infinite prediction horizon. Hence, it cannot be used for comparisons against the actual number of transactions. The same applies to its analog for finite prediction horizons, the discounted expected conditional transactions (DECT) metric. 

\begin{CodeChunk}
\begin{CodeInput}
R> mae <- function(x, y){return(mean(abs(x - y)))}
R> rmse <- function(x, y){sqrt(mean((x - y)^2))}
R> dt.pred[, list(
    +     mae.cet = mae(CET, actual.x),
    +     rmse.cet = rmse(CET, actual.x),
    +     mae.total.spending = 
    +         mae(CET * predicted.mean.spending, actual.total.spending),
    +     rmse.total.spending = 
    +         rmse(CET * predicted.mean.spending, actual.total.spending)
    + )]
\end{CodeInput}
\begin{CodeOutput}
     mae.cet   rmse.cet   mae.total.spending   rmse.total.spending
       <num>      <num>                <num>                 <num>
1:  2.039532   3.329395             87.64222                182.38
\end{CodeOutput}
\end{CodeChunk}

\subsubsection{Making final predictions without holdout period}

To obtain the final, most accurate estimates, we make predictions with a model that is estimated on all available purchasing data. As all models in \pkg{CLVTools} are generative models, no holdout period or any other form of holdout data is required to estimate them or to define a dependent variable.

In general, this is the final step in a workflow that in previous steps has mainly centered around model testing, e.g., by comparing different specifications in a holdout period. Once a user has decided on a final model configuration, the final model is estimated on all available transaction data. To this end, a new transaction data object without holdout period has to be created (\code{estimation.split = NULL}). It goes without saying that model fit should still be assessed by using diagnostics that do not require a holdout period. For example, a visual inspection of both tracking plot types. If the fit is found to be adequate, this model will then be used to make predictions for the focal customer cohort. 

When making predictions without a holdout period, the argument \code{prediction.end} must be given to specify how far into the future to predict. In this case, we predict 95 weeks into the future. For the DERT expression, a discount rate also needs to be specified. By default CLVTools uses 
\[
\delta = \ln(1 + 0.10),
\]
which corresponds to a 10\% discrete annual rate. The natural logarithm appears because continuous compounding models growth as \(e^{\delta}\); equating this to the discrete one-year growth factor \(1+d\) and solving for \(\delta\) gives \(\delta=\ln(1+d)\).%

If the \texttt{time.unit} chosen in \texttt{clvdata()} takes any value other than “yearly”, the continuous rate must be scaled down by the number of its occurrences per year. Specifically, with an annual discrete rate \(d\), the appropriate rate is computed as
\[
\delta_k = \frac{\ln(1 + d)}{k},
\]
where \(k\) is the number of time units per year (e.g.\ \(k = 52\) for weekly, \(k = 365\) for daily units). Hence, for a discrete annual rate of 7.5\%, and weekly units, one sets 
\[
\delta_{52} = \frac{\ln(1.075)}{52}.
\]
By passing \(\delta_k\) to \texttt{predict()}, we ensure that both the prediction horizon and the discounting procedure use time units in a consistent way with the data indexing defined through \texttt{time.units}.

\begin{CodeChunk}	
\begin{CodeInput}
R> clv.apparel.full <- clvdata(
+   data.transactions = apparelTrans,
+   date.format = "ymd", 
+   time.unit = "week",
+   estimation.split = NULL, # no holdout period
+   name.id = "Id", 
+   name.date = "Date",
+   name.price = "Price"
+ )
R> est.pnbd.full <- latentAttrition(family = pnbd, data = clv.apparel.full)
R> est.gg.full <- spending(family = gg, data = clv.apparel.full)
R> dt.pred.full <- predict(
+   est.pnbd.full,
+   predict.spending = est.gg.full, 
+   prediction.end = 95, 
+   continuous.discount.factor = log(1 + 0.075) / 52, 
+   verbose = FALSE
+ )
R> head(dt.pred.full, 3)
\end{CodeInput}
\begin{CodeOutput}
Key: <Id>
        Id   period.first   period.last   period.length        PAlive
    <char>         <Date>        <Date>           <int>         <num>
1:       1     2010-12-21    2012-10-15              95   0.007191623
2:      10     2010-12-21    2012-10-15              95   0.836860928
3:     100     2010-12-21    2012-10-15              95   0.922281780

           CET          DERT   predicted.mean.spending
         <num>         <num>                     <num>
1:  0.01300226    0.06200625                  77.79363
2:  0.89770449    4.28104733                  36.04491
3:  2.34558536   11.18582123                  37.23417

    predicted.period.spending   predicted.CLV
                       <num>           <num>
1:                  1.011493        4.823691
2:                 32.357674      154.309950
3:                 87.335919      416.494747
\end{CodeOutput}
\end{CodeChunk}

The metrics reported here, which were derived on the full purchasing data without any holdout period, are the final metrics that can be used for managerial decision making. For example, to select customers for a marketing campaign or as part of any marketing automation system. 

Recall that the prediction horizon (\code{prediction.end}) is only considered for metrics \code{CET} and \code{predicted.period.spending}, while the value of \code{continuous.discount.factor} only affects \code{DERT} and \code{predicted.CLV}. \code{PAlive} is unaffected by both parameters as it describes customers at the end of the estimation period. Due to their nature, \code{DERT} and \code{predicted.CLV} have an infinite time horizon. Thus, depending on the discount rate, they can be larger than the metrics that consider the given, finite prediction horizon. Method-specific deviations might be applied for models with time-varying covariates. See, e.g., the DECT metric, which relies on a finite prediction horizon and is detailed in Section \ref{chapter:advanced} \citep{Bachmann2021}. 

When executing this task on a regular basis (e.g., daily or weekly), practitioners may want to consider the following: The day after the model is estimated and predictions have been made, more data is available. However, parameters likely change only marginally. Thus, the previously fitted model can still be used to make predictions on the next day. To do so, the full dataset, which includes the new purchase records, can be given in \code{newdata}. The predictions are then made using the parameters of the already estimated model and using the input of the model (i.e., the purchase summary $(x, t_x, T)$) calculated on the order data in \code{newdata}.

\subsubsection{Making predictions that account for uncertainty}

Confidence intervals help users to quantify the uncertainty of a statistical model, i.e., they give an indication of the degree of uncertainty of an estimate. \pkg{CLVTools} facilitates the derivation of confidence intervals based on a bootstrapping routine.

To calculate confidence intervals for all predicted metrics, the \code{predict()} command accepts arguments \code{uncertainty} and \code{num.boots}. If the former argument is set to \code{"boots"}, the latter defines the number of bootstrap samples.
Customers, together with their entire purchasing history, are sampled with replacement. A new model is estimated on the sampled data with the same specification and optimization options as the given fitted model. 

For more custom applications, the \code{clv.bootstrapped.apply} command can be used. Given an estimated model, it samples new data from the transactions stored in it, re-fits the model on it, and then applies a user-given method on the newly estimated model. In each iteration, the customers together with their orders are selected by \code{fn.sample}. The model is fit on the new data with the exact same options as it was originally fit but any option can be overwritten by passing it in the ellipsis construct (\code{...}). The method \code{fn.boot.apply} is applied on the model fit on the bootstrapped data and its return value is collected in a list which is eventually returned. It should be noted that simply sampling customers with their orders and creating a data object may yield different estimation and holdout periods because the end of the data is determined by the last order. This method makes sure that the estimation and holdout periods are preserved as in the original data. In this way, the model inputs (i.e, $(x, t_x, T)$) in each iteration remain for each customer the same as in the original data, making it equivalent to sampling from the model inputs generated on the original data.

For example, it facilitates the following routine that visualizes this uncertainty within the tracking plot: 

\begin{CodeChunk}	
\begin{CodeInput}
R> set.seed(1)
R> l.boot <- clv.bootstrapped.apply(
+   object = est.pnbd, 
+   num.boot = 10, 
+   verbose = FALSE,
+   fn.sample = function(ids){
+     # method to sample customers each iteration
+     return(sample(ids, size = length(ids), replace = TRUE))}, 
+   fn.boot.apply = function(fitted){
+     # method applied on the model fitted on bootstrapped data
+     return(plot(
+       fitted, 
+       which = "tracking", 
+       plot = FALSE, # only return plot data
+       verbose = FALSE, 
+       transactions = FALSE # only model expectation, no actual num orders
+ ))})
R> dt.boot <- rbindlist(l.boot)
R> dt.CIs <- dt.boot[, list(
+   lb = quantile(value, 0.05), 
+   ub = quantile(value, 0.95)), 
+   by = "period.until"
+ ]
R> plot.original <- plot(est.pnbd, verbose = FALSE) 
R> plot.original + geom_ribbon(
+   aes(ymin = lb, ymax = ub), 
+   data = dt.CIs, 
+   alpha = 0.3, 
+   show.legend = FALSE
+ )
\end{CodeInput}
\end{CodeChunk}

\begin{figure}[h!]
    \centering
    \includegraphics[width=1\textwidth]{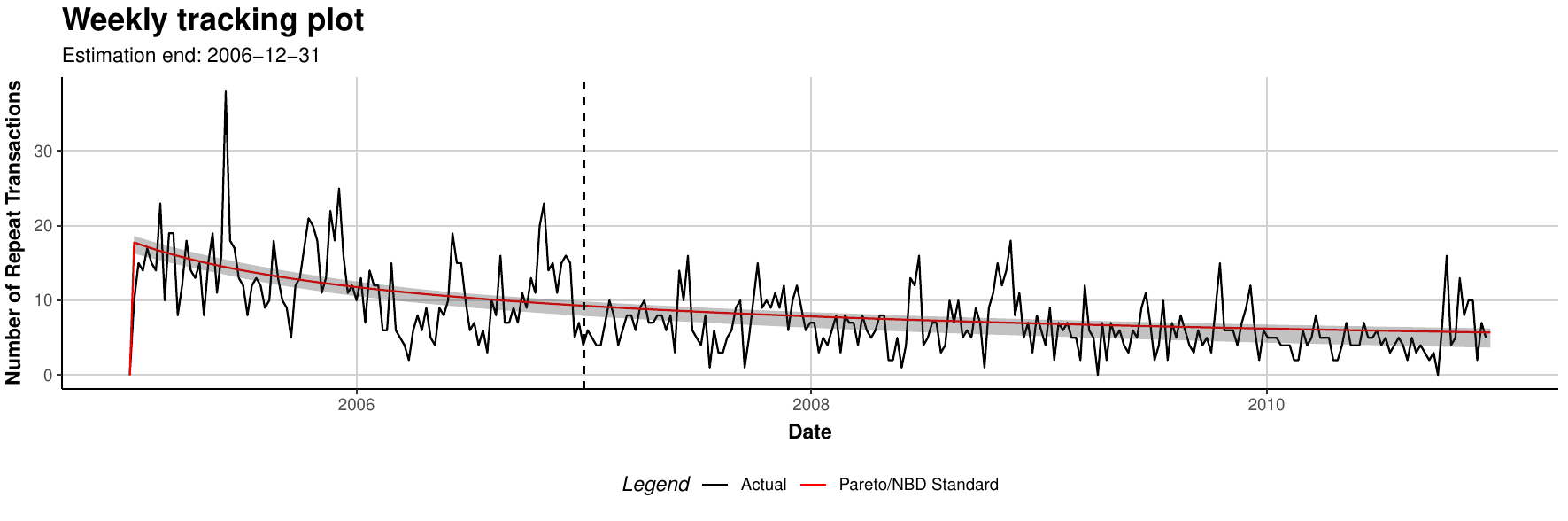}
    \caption{Tracking plot (non-cumulative) with confidence intervals}
    \label{plot:uncertainty}
\end{figure}

The resulting plot highlights the 5\% and 95\% confidence interval for the number of purchases of this customer cohort for the estimation and holdout period. It is shown in Figure \ref{plot:uncertainty}.

It should be noted that bootstrapping only accounts for uncertainty in model
parameters (epistemic uncertainty), and not sampling variability in the actual outcomes (aleatoric uncertainty). How to additionally consider the latter for this application case, i.e., the stochasticity in the data generating process, is an area of ongoing research.

\subsubsection{Making predictions for prospective customers}

While the discussed latent attrition models focus primarily on modeling the behavior of existing customers, they can also be used to derive the \textit{average} value of prospective customers that are  yet to be acquired.
To this end, the unconditional expectation used for the tracking plot can be leveraged as it gives the expected number of orders for a customer for which no information is available. 
The method \code{newcustomer()} builds on this to predict the total number of orders an average prospective customer will make in the first \code{t} periods. 
It should be emphasized that such predictions are for entirely hypothetical average customers with no order history. Therefore, none of the transaction data stored in the fitted model is used, nor can any order data be given to the \code{newcustomer()} method. The following expression, which is based on the Pareto/NBD model, highlights that no inputs derived from orders are required for this prediction:
\[
    1 + E[X(t)]= 1 + \frac{r \beta}{\alpha (s-1)} \left[  1- \left (\frac{\beta}{\beta+t} \right)^{s-1}  \right].
\]

We add +1 to the unconditional expectation to account for all transactions that a prospective customer will make, including the first one. Most applications of probabilistic models for customer base analysis focus on modeling repeat transactions that occur after the first transaction of a customer. In the following example, we predict the number of purchases of a prospective customer in the first year. To do so, the number of transactions and the average spending per transaction of a customer are separately predicted. To derive the expected total spending, we multiply both. 

\begin{CodeChunk}	
\begin{CodeInput}
R> est.gg.full.first <- spending(
+   family = gg, 
+   data = clv.apparel.full, 
+   remove.first.transaction=FALSE)
R> nc.trans <- predict(est.pnbd.full, newdata=newcustomer(num.periods = 52))
R> nc.spend <- predict(est.gg.full.first, newdata=newcustomer.spending())
R> cat("Average expected number of transactions:", nc.trans, "\n")
R> cat("Average expected spending per order:", nc.spend, "\n")
R> cat("Average spending in the first year:", nc.trans * nc.spend, "\n")
\end{CodeInput}
\begin{CodeOutput}
Average expected number of transactions: 2.218635
Average expected spending per order: 39.1372
Average spending in the first year: 86.83115
\end{CodeOutput}
\end{CodeChunk}

Note that the spending model should be fitted on all orders, including the initial purchases of each customer (\code{remove.first.transaction=FALSE}). The reason is that the predicted spending will be multiplied with the total number of transactions that also includes the initial order and not only repeat purchases.

When making these predictions for prospective customers, it is recommended to use the latent attrition and spending model for the latest acquisition cohort. It is likely that this cohort resembles  prospective customers the most. If covariates have been included in the model, it is possible to account for these and assess various scenarios. For example, to determine the difference between prospective customers in region A versus region B. For details, see the documentation of \code{newcustomer.static()} and \code{newcustomer.dynamic()}.

\subsection{Model specifications for covariate data}
\label{chapter:advanced}

Time-invariant and time-varying covariates can help to better model complex patterns in customers' purchase records. We can identify the added value of increasing model complexity by benchmarking such a model against a relevant baseline.

\subsubsection{Latent attrition models with covariates} 

\pkg{CLVTools} can account for the observed cross-sectional heterogeneity by including covariates. In latent attrition models, covariates affect the purchase process and the attrition process. It is possible to include different covariates for the two processes. In general, two types of covariates are distinguished, i.e., time-invariant and time-varying. The former include factors that do not change over time, such as information on customer demographics or customers' acquisition channel. The latter may change over time and include, for example, marketing activities or seasonal patterns.

Data for time-invariant covariates must contain a unique customer identifier and a single value for each covariate. In the case of the apparel retailer, customers' gender and information on the acquisition channel are available as time-invariant covariates. Use the \code{data("apparelStaticCov")} command to load the time-invariant covariates. In this example, \code{gender} is coded as a dummy variable with \code{male=0} and \code{female=1} and \code{channel} with \code{online=0} and \code{offline=1}. 

\begin{CodeChunk}
\begin{CodeInput}
R> data("apparelStaticCov")
R> head(apparelStaticCov, n = 3)
\end{CodeInput}
\begin{CodeOutput}
        Id   Gender   Channel
    <char>    <num>     <num>
1:       1        0         0
2:       2        1         0
3:       3        1         0
\end{CodeOutput}
\end{CodeChunk}

Data for time-varying covariates require a time series of covariate values for every customer. In other words, if a time-varying covariate is included and the analysis is done based on weekly data, the covariate value can change every week. Thus, a value has to be specified for every customer every week. Note that the time unit specified for the transaction data also defines the time unit for the covariate data.

In our case study, data on seasonal patterns (\code{High.Season}) is available as a time-varying covariate. It indicates whether a week falls into the high season or not. Additionally, we have the gender and the sales channel that a customer used for the first purchase as a time-invariant covariate. The data structure of time-invariant covariates (i.e., \code{Gender} and \code{Channel}) needs to be aligned with the structure of time-varying covariates (i.e., \code{High.Season}). In other words, the former has to be repeated for every entry of the latter. Indeed, such a representation is common for time-series analyses that allow for both time-invariant and -varying covariates. 

\begin{CodeChunk}
\begin{CodeInput}
R> data("apparelDynCov")
R> head(apparelDynCov, n = 3)
\end{CodeInput}
\begin{CodeOutput}
        Id     Cov.Date   High.Season   Gender   Channel
    <char>       <Date>         <num>    <num>     <num>
1:       1   2005-01-02             0        0         0
2:       1   2005-01-09             0        0         0
3:       1   2005-01-16             0        0         0
\end{CodeOutput}
\end{CodeChunk}

In a next step, we need to add the covariate data to the transaction data object. Thereby, two commands are available:  \code{SetStaticCovariates()} and \code{SetDynamicCovariates()}. Both commands are mutually exclusive. The arguments \code{data.cov.life} and \code{data.cov.trans} are the \code{data.frame} or \code{data.table} that contain the covariate data for the attrition and the transaction process, respectively. If a covariate can affect both processes it has to be added in both arguments: \code{data.cov.life} \textit{and} \code{data.cov.trans}. Categorical data (\code{factor} and \code{character}) is turned into k-1 dummy variables. \code{names.cov.life} and \code{names.cov.trans} specify the columns in the given covariate data that should be added to the purchasing data. 

\begin{CodeChunk}
\begin{CodeInput}
R> clv.static <- SetStaticCovariates(
+   clv.data = clv.apparel, 
+   data.cov.life = apparelStaticCov, 
+   data.cov.trans = apparelStaticCov,
+   names.cov.life = c("Gender", "Channel"), 
+   names.cov.trans = c("Gender", "Channel"), 
+   name.id = "Id"
+ )
\end{CodeInput}
\end{CodeChunk}

For the second example, we have mixed types of covariates and use the following command:

\begin{CodeChunk}
\begin{CodeInput}
R> clv.mixed <- SetDynamicCovariates(
+   clv.data = clv.apparel, 
+   data.cov.life = apparelDynCov,
+   data.cov.trans = apparelDynCov, 
+   names.cov.life = c("High.Season", "Gender", "Channel"), 
+   names.cov.trans = c("High.Season", "Gender", "Channel"), 
+   name.id = "Id", 
+   name.date = "Cov.Date"
+ )
\end{CodeInput}
\end{CodeChunk}

Once the covariates are added to the transaction data object, the estimation process is almost identical to the one for a model without covariates. The same \code{latentAttrition()} interface is used again and a specific model implementation is selected based on the data passed in. For example, for the Pareto/NBD model, an optimized implementation is available depending on (a) whether or not the data includes covariates and (b) whether these covariates are time-invariant or time-varying. 

If a model is estimated with time-invariant or -varying covariates, the parameter \code{formula} must be given. The formula interface is used to select and transform the covariates stored in the data object before the model is estimated on it. It has to be a two-part formula without a dependent variable. The left-hand side of the formula has to remain empty as there is no dependent variable in the classical sense (see Section \ref{probmodels} for details). All required model inputs except the covariates are entirely predefined and are derived from the transaction data (i.e. $(x_i, t_{x_i}, T_i)$). The right-hand side follows a two-part notation using \code{|} as a separator. The first part specifies which covariates to include for the \textit{attrition process} while the second part specifies which covariates to include for the \textit{purchase process}. This allows each process to further transform the data or to add interactions. 

Further, start values for optimization can be defined for the covariate parameters using the arguments \code{start.params.life} and \code{start.params.trans}. If not given, the start values are set to 0.1 for all covariates.

We first show how the Pareto/NBD model with extension for time-invariant covariates is estimated. In addition to the model family, the parameter \code{formula} now has to be given to specify which of the covariates available in the data object should be used. The usual formula notation applies: Covariates can be specified by name or, if the aim is to include all covariates available for the respective process, by the placeholder \code{.} . In the following, we select all of the available covariates for both processes by explicitly naming them. To inspect the estimated model, we use \code{summary()}.

\begin{CodeChunk}
\begin{CodeInput}
R> est.pnbd.static <- latentAttrition(
+   formula = ~ Gender + Channel | Gender + Channel,
+   family = pnbd, 
+   data = clv.static
+ )
R> summary(est.pnbd.static)
\end{CodeInput}
\begin{CodeOutput}
Pareto/NBD with Static Covariates Model 

Call:
latentAttrition(formula = ~ Gender + Channel | Gender + Channel, 
family = pnbd, data = clv.static)

Fitting period:                                
Estimation start  2005-01-02    
Estimation end    2006-12-31    
Estimation length 104.0000 Weeks

Coefficients:
Estimate Std. Error  z-val Pr(>|z|)    
r               1.8378     0.3455     NA       NA 
alpha          92.9123    16.9670     NA       NA 
s               0.5920     0.2609     NA       NA   
beta           49.6227    36.2509     NA       NA    
life.Gender    -0.6430     0.2955 -2.176  0.02957 *  
life.Channel    0.7907     0.3059  2.585  0.00973 ** 
trans.Gender    0.2859     0.1041  2.745  0.00605 ** 
trans.Channel   0.6241     0.1050  5.946 2.74e-09 ***
---
Signif. codes:  0 ‘***’ 0.001 ‘**’ 0.01 ‘*’ 0.05 ‘.’ 0.1 ‘ ’ 1

Optimization info:                 
LL     -5821.0627
AIC    11658.1254
BIC    11693.3009
KKT 1  TRUE      
KKT 2  TRUE      
fevals 41.0000   
Method L-BFGS-B  

Used Options:                     
Correlation     FALSE
Regularization  FALSE
Constraint covs FALSE
\end{CodeOutput}
\end{CodeChunk}

The covariate parameters are directly interpretable as rate elasticities: a 1\% change in a covariate $\mathbf{x}^{P}$ or $\mathbf{x}^{A}$ changes the average purchase or the attrition rate by $\gamma_{purch}\mathbf{x}^{P}$ or $\gamma_{life}\mathbf{x}^{A}$ percent, respectively \citep{Gupta1991}. When dummy variables are used as covariates, the interpretation is relative to the baseline, i.e., the state defined as 0. 
In the case of the apparel retailer, we observe that female customers have a significantly higher purchase rate (\code{trans.Gender = 0.2859}) which, all other things equal, results in more purchases on average. Note that female customers are coded as 1, and male customers as 0. Also, customers acquired offline, coded as 1, purchase more (\code{trans.Channel = 0.6241}) but drop out more quickly (\code{life.Channel = 0.7907}) than customers who have been acquired online. 

Note that the four Pareto/NBD base parameters (\(r,\ \alpha,\ s,\ \beta\)) are not reported with any \(z\)- and \(p\)-values. As these parameters are constrained to be strictly positive, the model definition fixes their lower bound at 0. Thus, a null hypothesis of \(\theta = 0\) lies outside the admissible parameter space. Reporting ``\texttt{NA}'' for the corresponding \(z\)- and \(p\)-values therefore forestalls the boundary‐parameter inference problem and spares users the misleading suggestion that a base parameter judged ``insignificant'' should be set to zero.

To estimate the Pareto/NBD model with time-varying and time-invariant covariates, model estimation is analogous. What differs is only the transaction object passed in and the parameter \code{formula} which has to be adapted to include the additional covariates. Note that the model estimation with time-varying covariates is computationally much more demanding than the previously detailed alternatives. It is recommended to keep an eye on the progress of model optimization. If necessary, the calculation of the Hessian matrix can be skipped to decrease runtime. The variance-covariance matrix is then not available, including statistics based on it such as standard deviation and p-values.  To do so, a user can specify \code{optimx.args}: 
\begin{CodeChunk}
\begin{CodeInput}
R> est.pnbd.mixed <- latentAttrition(
+   formula = 
+       ~ Gender + Channel + High.Season | Gender + Channel + High.Season, 
+   family = pnbd, 
+   data = clv.mixed, 
+   # Print progress and disable calculation of the Hessian matrix.
+   # The latter requires to also disable calculating KKT.
+   optimx.args = list(hessian=FALSE, control = list(trace = 1, kkt=FALSE))
+ )
\end{CodeInput}
\end{CodeChunk}

If the Hessian matrix is needed at a later point, it can be obtained using the method \code{hessian()}. By adding it to the fitted model object, all functionalities that rely on it become available, such as standard errors. 
See the vignette with advanced techniques for illustrations.

Note that in contrast to a Pareto/NBD model that does not include any covariates or only time-invariant covariates, the default optimizing routine changes. For a Pareto/NBD model with time-varying covariates, we use as a default the robust
but often slower Nelder-Mead method (Nelder and Mead 1965). Users can always override this default and choose an alternative optimizing routine such as L-BFGS-B (Byrd,
Lu, Nocedal, and Zhu 1995).

To predict future purchase behavior, we call \code{predict()} on the estimated model. For the case of time-varying covariates, covariates must be provided for the entire prediction period. The covariates stored in the transaction data object, on which the model was estimated, may not cover the desired prediction period. In this case, a new order data object with long enough covariate data can be created and supplied as argument \code{newdata} (see Section \ref{chapter:FinalPredictionsWithDyncoc}).

Note that the predicted metrics slightly differ for models with time-varying covariates. Instead of DERT, DECT is predicted which only covers a finite time horizon in contrast to DERT (see \citet{Bachmann2021} for details).

To decide on a model configuration, diagnostic plots from multiple models can be visualized in a single plot to facilitate comparison. Here, we contrast the performance of the three previously discussed models: (1) the base model without any covariates (\code{est.pnbd}), (2) the model with time-invariant covariates (\code{est.pnbd.static}), and (3) the model with both time-invariant and time-varying covariates (\code{est.pnbd.mixed}). 

Here, we focus on the tracking plot as an example. To start with, we learn from the non-cumulative plot (\code{cumulative=FALSE}) that the model with time-varying covariates indeed fits the seasonal spikes much better than the alternative models that do not account for such patterns. From the cumulative tracking plot (\code{cumulative=TRUE}), we learn that the model with time-invariant and time-varying covariates also seems to have the best aggregate predictions. While the former plot helps to visually inspect efforts in capturing the seasonal patterns, the latter plot is often more intuitive for assessing the models' relative performance. Note that for both tracking plot types, an assessment is possible for in-sample and out-of-sample observations. 

\begin{figure}[h!]
    \centering
    \vspace{0.3cm}	\includegraphics[width=1\textwidth]{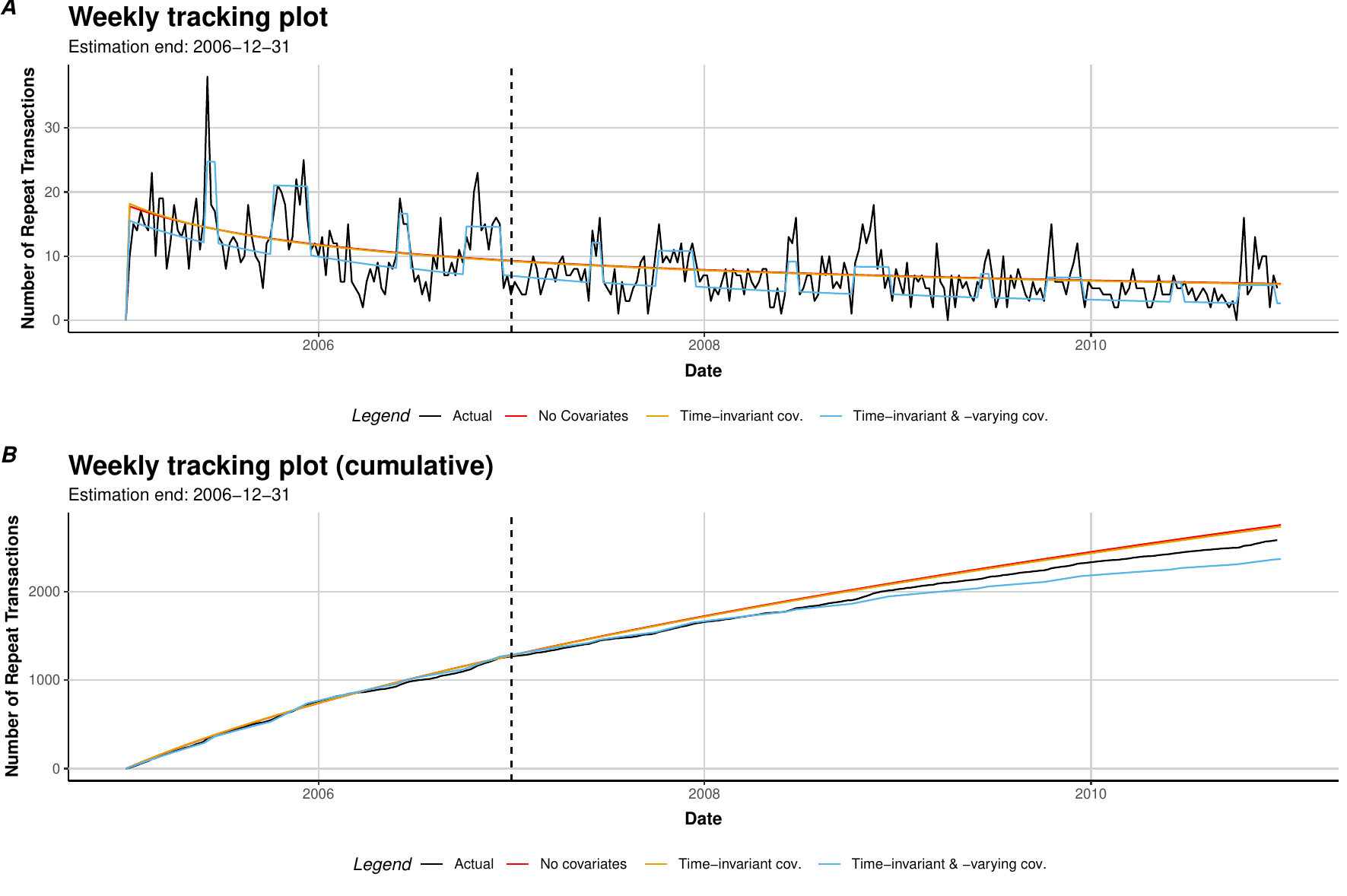}
    \caption{Tracking plot for the apparel dataset - model comparison (A: n.-cumul.; B: cumul.)}
    \label{plot:tracking-model-comparison}
\end{figure}

The non-cumulative and cumulative plots for our case study are shown in Figure \ref{plot:tracking-model-comparison}. 

\begin{CodeChunk}
\begin{CodeInput}
R> plot.12.labels <- c(
+   "No Covariates", 
+   "Time-invariant cov.",
+   "Time-invariant & -varying cov."
+ )
R> plot.12a <- plot(
+   est.pnbd,
+   which = "tracking",
+   cumulative = FALSE,
+   other.models = list(est.pnbd.static, est.pnbd.mixed),
+   label = plot.12.labels
+ )
R> plot.12b <- plot(
+   est.pnbd,
+   which = "tracking", 
+   cumulative = TRUE, 
+   other.models = list(est.pnbd.static, est.pnbd.mixed),
+   label = plot.12.labels
+ )
R> ggarrange(plot.12a, plot.12b, labels = c("A", "B"), ncol = 1, nrow = 2)
\end{CodeInput}
\end{CodeChunk}

Out-of-sample performance can also be measured with widely known metrics. Thereby, we rely on the actual values for the holdout period. For convenience, these are also included in the output from \code{predict}. Note that in this case, \code{prediction.end} defaults to the end of the holdout period.

\begin{CodeChunk}
\begin{CodeInput}
R> data.frame(
+   pnbd.std = t(
+     predict(est.pnbd)[,list(
+       MAE = mae(CET, actual.x), 
+       RMSE = rmse(CET, actual.x))]),
+   pnbd.static = t(predict(est.pnbd.static)[, list(
+     MAE = mae(CET, actual.x),
+     RMSE = rmse(CET, actual.x))]),
+   pnbd.mixed = t(predict(est.pnbd.mixed)[, list(
+     MAE = mae(CET, actual.x),
+     RMSE = rmse(CET, actual.x))])
+ )
\end{CodeInput}
\begin{CodeOutput}
        pnbd.std   pnbd.static   pnbd.mixed
MAE     2.039532      2.006879   1.857776
RMSE    3.329395      3.276752   3.253590
\end{CodeOutput}
\end{CodeChunk}

For further model assessment, it is possible to look at the AIC and BIC values reported by the \code{summary()} command or alternatively, the generics \code{AIC()} and \code{BIC()}. For our application case, the results support the previous findings. They indicate that the model including time-invariant and time-varying covariates approximates the underlying data-generating process best.

\subsubsection{Making final predictions without holdout period (with time-varying covariates)}
\label{chapter:FinalPredictionsWithDyncoc}

In section \ref{chapter:prediction}, we discussed how to derive the final predictions after the model was estimated on the all available transaction data without a holdout period. The motivation for this step is to use all available data to derive the customer-level predictions that are used in production. However, this task of making predictions beyond the last recorded purchase is more complex when time-varying covariates are included in a model. The reason is that the future state of all time-varying covariates has to be provided by the user for the entire prediction period. This is also the reason as to why only a few time-varying covariates are generally available for predictive modeling of time series: In most cases, the state of time-varying variables is not known in the future. An exception is seasonality patterns that depend on known dates such as school holidays or Black Friday. 

For our case study, we illustrate how to make predictions based on the model configuration that was found to be best suited during model benchmarking. Thereby, we re-define the input data without the holdout period and then re-estimate the latent attrition and spending model.  

\begin{CodeChunk}
\begin{CodeInput}
R> clv.apparel.full <- clvdata(
+   data.transactions = apparelTrans, 
+   date.format = "ymd", 
+   time.unit = "week",
+   estimation.split = NULL, # no holdout period
+   name.id = "Id",
+   name.date = "Date",
+   name.price = "Price"
+ )
R> clv.apparel.mixed.full <- SetDynamicCovariates(
+   clv.data = clv.apparel.full, 
+   data.cov.life = apparelDynCov,
+   data.cov.trans = apparelDynCov, 
+   names.cov.life = c("High.Season", "Gender", "Channel"), 
+   names.cov.trans = c("High.Season", "Gender", "Channel"), 
+   name.id = "Id", 
+   name.date = "Cov.Date"
+ )
R> est.pnbd.mixed.full <- latentAttrition(
+   formula = 
+       ~ Gender + Channel + High.Season | Gender + Channel + High.Season,  
+   family = pnbd,
+   data = clv.apparel.mixed.full,
+   optimx.args = list(hessian=FALSE, control = list(trace = 1, kkt=FALSE))
+ )
R> est.gg.full <- spending(family = gg, data = clv.apparel.mixed.full)
\end{CodeInput}
\begin{CodeOutput}
\end{CodeOutput}
\end{CodeChunk}

Next, we derive the final predictions. However, the information on the time-varying covariates, which was stored in \code{clv.apparel.mixed.full} and used to estimate the model, covers only the time during which purchases were recorded. To predict beyond this time point, we have to provide covariate information which reaches into the future, i.e.,  covers the prediction period. 

For our case study, we have prepared this data accordingly. It covers nearly two years beyond the last purchase. We load this information by executing \code{data(apparelDynCovFuture)},:

\begin{CodeChunk}
\begin{CodeInput}
R> data("apparelDynCovFuture")
R> cat("Range of covariates for estimation: \n",
+   format(range(apparelDynCov$Cov.Date)), "\n")
R> cat("Range of covariates for prediction: \n", 
+   format(range(apparelDynCovFuture$Cov.Date)), "\n")
\end{CodeInput}
\begin{CodeOutput}
Range of covariates for estimation
 2005-01-02 2010-12-26
Range of covariates for prediction"
 2011-01-02 2012-10-14
\end{CodeOutput}
\end{CodeChunk}

In \code{data(apparelDynCovFuture)}, we have extended the covariate data with the known future values of the time-varying and time-invariant covariates in the prediction period. To use this information about the future states of the covariates to make predictions, we have to combine it with the covariate information for the estimation period, i.e., \code{data(apparelDynCov)}. The reason for this is that some model expressions are derived from the covariate information of the estimation period. Next, these covariate data are added to the transaction data object.  

It should be highlighted that there is no need to re-estimate the model at this point. Instead, we can execute the \code{predict()} method on the already estimated model and pass the transaction data object with extended covariates as parameter \code{newdata}.

\begin{CodeChunk}
\begin{CodeInput}
R> apparelDynCovPastAndFuture <- rbind(apparelDynCov, apparelDynCovFuture)
R> clv.apparel.mixed.PastAndFuture <- SetDynamicCovariates(
+   clv.data = clv.apparel.full, 
+   data.cov.life = apparelDynCovPastAndFuture,
+   data.cov.trans = apparelDynCovPastAndFuture, 
+   names.cov.life = c("High.Season", "Gender", "Channel"), 
+   names.cov.trans = c("High.Season", "Gender", "Channel"), 
+   name.id = "Id", 
+   name.date = "Cov.Date"
+ )
R> dt.pred.mixed.future <- predict(
+   est.pnbd.mixed.full,
+   newdata=clv.apparel.mixed.PastAndFuture,
+   predict.spending = est.gg.full,
+   prediction.end = 95, 
+   continuous.discount.factor = log(1 + 0.075) / 52
+ )
R> head(dt.pred.mixed.future, 3)
\end{CodeInput}
\begin{CodeOutput}
Key: <Id>
        Id   period.first   period.last   period.length      PAlive
    <char>         <Date>        <Date>           <int>       <num>
1:      1     2010-12-21    2012-10-15              95   0.0139206
2:     10     2010-12-21    2012-10-15              95   0.8108995
3:    100     2010-12-21    2012-10-15              95   0.9103230

          CET        DECT   predicted.mean.spending
        <num>       <num>                     <num>
1:  0.0253848   0.02379146                 77.79363
2:  1.5938786   1.49351918                 36.04491
3:  4.0419238   3.78742184                 37.23417

    predicted.period.spending   predicted.period.CLV
                        <num>                  <num>
1:                   1.974776               1.850824
2:                  57.451205              53.833759
3:                 150.497668             141.021501
\end{CodeOutput}
\end{CodeChunk}

An extension to include time-invariant and time-varying covariates in the GG model -- following the approach proposed by \citet{Bachmann2022a, Bachmann2022b} -- will be added in the future to \pkg{CLVTools}.

\subsection{Advanced modeling techniques}\label{chapter:explaining} 

Going beyond conducting a basic customer base analysis with or without covariates, various advanced modeling options are provided in \pkg{CLVTools}. These include the ability (1) to add regularization for covariate parameters, (2) to account for the correlation between the transaction and dropout process, and (3) to set equality constraints on covariate parameters.    

An additional use case for advanced modeling techniques is to control for endogenous covariates. The covariate parameter estimates for the covariates can give some insight into what drives customers' purchase behavior. If the exogeneity assumption of the covariates is violated, various techniques can be used to control for this. Thereby, the models that support covariates in CLVTools can be used together with other packages that implement related two-step modeling techniques. A first option is to use instrumental variables. If these are not available, internal instrumental variable approaches can serve as an alternative \citep{Gui2023}. For a case study detailing various approaches to controlling the endogeneity of marketing campaigns, see \citet{Bachmann2021}. In our case study, all covariates are assumed to be exogenous. 

For further advanced techniques, see \citet{Meierer2025a}.

\subsubsection{Regularization of covariate parameters}	

When a large number of covariates are included in the analysis, regularization can help prevent overfitting. To this end, it is possible to apply a normal prior on the covariate parameters (L2 regularization). This requires specifying a regularization weight per process. The value of the regularization weight is the same for all covariate parameters of a process. The larger this regularization weight, the stronger the effect of the regularization. To find the optimal value, any hyperparameter optimization procedure can be applied. 

To regularize covariate parameters, the regularization weights for both processes have to be defined in the parameter \code{reg.lambdas} in \code{latentAttrition()}. 
For example, by specifying \code{reg.lambdas = c(trans = 0.1, life = 0.1)}, a user sets the regularization weight to 0.1 for both processes. The use of regularization is indicated at the end of the output of \code{summary()}. 

Given the limited number of covariates in our case study, adding regularization of their parameter estimates is not necessary.

\subsubsection{Adding a correlation between the transaction and dropout process}

To relax the assumption of independence between the transaction and the attrition process, specify the argument \code{use.cor} in the \code{latentAttrition()} command. This is independent of whether the model includes covariates or not. With regards to the latter, this is an extension of the model presented in \citet{Bachmann2021}. In the case of \code{use.cor=TRUE}, a Sarmanov approach is used to correlate the attrition and transaction process (see section \ref{advanced-modeling} for details). The argument \code{start.param.cor} allows us to optionally specify a starting value for the correlation parameter.

The model output will then list an additional parameter \code{Cor(life,trans)}, which corresponds to $p_m$ in \eqref{eq:corrkoefficient} and can be directly interpreted as a correlation:
\begin{itemize}
    \item If the correlation is zero, it indicates that there is no relationship between customers’ transaction and attrition rate. 
    \item If the correlation is positive and significant, customers with a higher (lower) transaction rate are more (less) likely to churn. The underlying mechanism is as follows:  a higher transaction rate $\lambda$ is associated with a higher attrition rate $\mu$, i.e., a reduction in the customer's lifetime.
    \item If the correlation is negative and significant, customers with a higher (lower) transaction rate are less (more) likely to churn.     
\end{itemize}

The impact of adding a correlation parameter depends on the dataset. In many applications that focus on prediction rather than on an in-depth understanding of customers' purchase behavior, the modeling of the additional parameter is neglected. A key reason for this is the increase in computational complexity compared to the often only marginal change in predictive accuracy. While this is a common decision among practitioners, it depends on the data at hand and the modeling objective. 

For this case study, adding this correlation does indeed have a limited impact on predictive accuracy. This applies to all previously tested models. Therefore, we proceed without considering this correlation.  

\subsubsection{Including equality constraints for covariate parameters}

If more complex hypothesis testing is required, users can leverage parameter constraints to compare effect sizes between the attrition and transaction process. All latent attrition models that can account for time-invariant and -varying covariates support equality constraints for the respective covariate parameters. For example, it is possible to test whether the parameter value of the covariate \code{gender} is the same for both processes. This facilitates testing of novel hypotheses and thus, helps to increase the understanding how covariates impact each process.

For our case study data, we add such a constraint (\code{constr.Gender}): 

\begin{CodeChunk}
\begin{CodeInput}
R> est.pnbd.mixed.constr <- latentAttrition(
+   formula = ~ . | ., 
+   names.cov.constr = "Gender", 
+   family=pnbd, 
+   data = clv.apparel.mixed.full,
+   optimx.args = list(control = list(trace = 1))
+ )
R> summary(est.pnbd.mixed.constr)
\end{CodeInput}
\begin{CodeOutput}
Pareto/NBD with Dynamic Covariates  Model 

Call:
latentAttrition(formula = ~. | ., family = pnbd, 
    data = clv.apparel.mixed.full, optimx.args = list(
    control = list(trace = 1)), names.cov.constr = "Gender")

Fitting period:                                
Estimation start  2005-01-02    
Estimation end    2010-12-20    
Estimation length 311.1429 Weeks

Coefficients:
Estimate Std. Error  z-val Pr(>|z|)    
r                  1.68794    0.21789     NA       NA 
alpha             93.02150   13.82645     NA       NA 
s                  0.50273    0.07909     NA       NA 
beta              55.14172   16.48224     NA       NA 
life.High.Season   0.54747    0.55919  0.979 0.327560    
life.Channel       0.69076    0.22692  3.044 0.002334 ** 
trans.High.Season  0.68975    0.04083 16.894  < 2e-16 ***
trans.Channel      0.57020    0.09444  6.038 1.56e-09 ***
constr.Gender      0.14102    0.09248  1.525 0.127273    
---
Signif. codes:  0 ‘***’ 0.001 ‘**’ 0.01 ‘*’ 0.05 ‘.’ 0.1 ‘ ’ 1

Optimization info:                  
LL     -11775.9091
AIC    23569.8182 
BIC    23609.3906 
KKT 1  TRUE       
KKT 2  TRUE       
fevals 1826.0000  
Method Nelder-Mead

Used Options:                           
Correlation          FALSE 
Regularization       FALSE 
Constraint covs      TRUE  
Constraint params    Gender
\end{CodeOutput}
\end{CodeChunk}

Here, an additional model is estimated that forces the covariate \code{gender} to have the same parameter value for the transaction and attrition process. We use the argument \code{names.cov.constr} to specify this variable. In consequence, the model output only contains a single parameter value for \code{gender}. The use of parameter constraints is indicated in the \code{summary()} output. A likelihood ratio test helps to evaluate if adding an equality constraint changes the model fit.

\begin{CodeChunk}
\begin{CodeInput}
R> lrtest(
+   est.pnbd.mixed.constr, 
+   est.pnbd.mixed.full, 
+   name = c("Constrained Model", "Unconstrained Model")
+ )
\end{CodeInput}
\begin{CodeOutput}
Likelihood ratio test

Model 1: Constrained Model
Model 2: Unconstrained Model
#Df LogLik Df  Chisq Pr(>Chisq)    
1   9 -11776                         
2  10 -11763  1 26.072  3.289e-07 ***
---
Signif. codes:  0 ‘***’ 0.001 ‘**’ 0.01 ‘*’ 0.05 ‘.’ 0.1 ‘ ’ 1
\end{CodeOutput}
\end{CodeChunk}

By comparing the likelihood values of the unconstrained and the constrained model, the test results indicate whether the parameter of a covariate significantly differs between the attrition and transaction process. In our case study, the results show a significant difference for \code{gender}. Thus, we conclude that the parameter of the constrained variable for the attrition and transaction process is statistically significantly different. In other words, the model fit has worsened significantly by adding an equality constraint for the parameter value of \code{gender}. If exogenous information on marketing interventions is available, such an analysis can be helpful to disentangle the way that a marketing intervention impacts customers' purchase behavior.

\section{Conclusion}

With \pkg{CLVTools}, we provide a comprehensive implementation framework for probabilistic models to predict the future purchase behavior of individual customers. The package features a standardized workflow independent of the specific model. \pkg{CLVTools} is based on a flexible class-based framework enabling the accommodation of various model variations. Additionally, the package is the first to provide options for advanced modeling techniques such as including time-invariant and time-varying covariates, correlation between the lifetime and transaction processes, regularization, and equality constraints. 





\bibliography{jss5634}


\newpage

\begin{appendix}
    

    \section{Marginalized Likelihood of the Pareto/NBD model} \label{app:pnbd_ll}
    
    \begin{align}\label{eq:standardlikelihood}
          & L(r, \alpha, s, \beta | x, t_x, T) \nonumber                                                                                                                                                                              \\
          & = \int_{0}^{\infty} \int_{0}^{\infty} \left( \frac{\lambda^x \mu}{\lambda + \mu} e^{-(\lambda + \mu) t_x} + \frac{\lambda^{x+1} \mu}{\lambda + \mu} e^{-(\lambda + \mu) T} \right) g(\lambda| r, \alpha) g(\mu) \nonumber \\
          & =\frac{\Gamma(r + x) \alpha^r \beta^s}{\Gamma(r)} \left( \frac{s}{r+s+x}  A_1 + \frac{r+x}{r+s+x} A_2 \right),                                                                                                            
    \end{align}
    with 
    \begin{align*}
        A_1= \begin{cases} \frac{_2 F_1(r+s+x, s+1, r+s+x+1; \frac{\alpha-\beta}{\alpha + t_x})}{(\alpha + t_x)^{r+s+x}} & \text{ if } \alpha \geq \beta          \\
        \frac{_2 F_1(r+s+x, r+x, r+s+x+1; \frac{\beta-\alpha}{\beta + t_x})}{(\beta + t_x)^{r+s+x}}                      & \text{ if } \alpha < \beta \end{cases} 
    \end{align*}
    and
    \begin{align*}
        A_2=\begin{cases} \frac{_2 F_1(r+s+x, s, r+s+x+1; \frac{\alpha-\beta}{\alpha + T})}{(\alpha + T)^{r+s+x}} & \text{ if } \alpha \geq \beta           \\
        \frac{_2 F_1(r+s+x, r+x + 1, r+s+x+1; \frac{\beta-\alpha}{\beta + T})}{(\beta + T)^{r+s+x}}               & \text{ if } \alpha < \beta \end{cases}. 
    \end{align*}
    The symbol $_2 F_1(a,b,c,z)$ refers to the integral representation of the Gaussian hypergeometric function \citep[see e.g.][]{abramowitz1964}. 
    
\newpage

 \section{Comparison of Model Run Times} \label{app:run_times}

As discussed in Section \ref{guidance}, we compare the run times of three variants of the Pareto/NBD model under varying sample sizes and estimation period lengths (in months). The results, showing average runtime in minutes, are summarized in Table~\ref{tab:benchmarking}. All models are estimated based on simulated data. We used the following base parameters for the Pareto/NBD model for this: $r = \text{1}, \alpha = \text{0.5}, s = \text{1}, \beta = \text{0.5}$ . Each scenario has been tested 3 times and the median run time (in minutes) is reported. We look at 3 model variations which differ in their estimation period and sample size: (1) a model without any covariate (2) a model with a single, dummy-coded time-invariant covariate, and (3) a model with a single, dummy-coded time-varying covariate. The latter two models illustrate the use of variables such as gender or seasonality indicators. In all cases, we use the optimization method \code{L-BFGS-B} \citep{byrd1995limited}. All simulations were performed without calculating the Hessian. These analyses were run on a consumer-grade computer (AMD Ryzen 9 9950X 16-Core Processor). All algorithms were using a single core to ensure a high degree of comparability.

\renewcommand{\arraystretch}{1.2}
\begin{table}[htbp]
    \centering
    \small
    \caption{Run time comparison for the Pareto/NBD Model for various scenarios}
    \label{tab:benchmarking}
    \begin{tabular}{
        l 
        >{\centering\arraybackslash}p{2cm} 
        >{\raggedleft\arraybackslash}p{3cm} 
        >{\raggedleft\arraybackslash}p{3cm} 
        >{\raggedleft\arraybackslash}p{3cm} 
        }
        \toprule
        \multirow{3}{*}{\shortstack[l]{\textbf{Sample} \\ \textbf{Size} \\ \textbf{(Customers)}}}
        & \multirow{3}{*}{\shortstack[c]{\textbf{Estimation} \\ \textbf{Period} \\ \textbf{(Weeks)}}}
        & \multicolumn{3}{c}{\textbf{Models}} \\
        \cmidrule(l{.5em}r{.5em}){3-5}
        & 
        & \multicolumn{1}{r}{\shortstack[r]{\textbf{No} \\ \textbf{Covariate}}}
        & \shortstack[r]{\textbf{Time-Invariant} \\ \textbf{Covariate}}
        & \shortstack[r]{\textbf{Time-Varying} \\ \textbf{Covariate}} \\
        \midrule
        1,000     & 26   & 0.19 sec & 0.28 sec & 6.28 sec \\
        1,000     & 52 & 0.19 sec & 0.26 sec & 2.15 sec \\
        1,000     & 104 & 0.20 sec & 0.26 sec & 10.68 sec \\
        \midrule
        10,000    & 26  & 0.27 sec & 0.55 sec & 8.18 sec \\
        10,000    & 52 & 0.32 sec & 0.57 sec & 21.37 sec \\
        10,000    & 104 & 0.37 sec & 0.72 sec & 58.42 sec \\
        \midrule
        100,000   & 26  & 0.43 sec & 1.02 sec & 1 min 30 sec \\
        100,000   & 52 & 0.62 sec & 1.34 sec & 3 min 16 sec \\
        100,000   & 104 & 1.01 sec & 1.49 sec & 10 min 39 sec \\
        \midrule
        1,000,000 & 26  & 1.23 sec & 2.07 sec & 12 min 34 sec \\
        1,000,000 & 52 & 1.58 sec & 3.43 sec & 34 min 22 sec \\
        1,000,000 & 104 & 2.33 sec & 5.60 sec & 134 min 40 sec \\
        \bottomrule
    \end{tabular}
\end{table}

Using the apparel and CDNOW dataset \citep{BTYD}, further analyses show that 
\begin{itemize}
\item generally the run time can be reduced by approximately 20-30 \% if the calculation of the Hessian is deactivated. This can be helpful in situations where the information of the Hessian is not required as prediction is the primary focus of the analysis at hand, and  
\item adding correlation to the Pareto/NBD model with the Sarmanov approach increases the computational time by a factor of approximately 2.5 to 6. 
\end{itemize}

Note that these results are only valid for the current implementation of the Pareto/NBD model in \pkg{CLVTools} and the hardware used. Using the optionally available parallelization techniques ,e.g., for the Extended Pareto/NBD model would lead to speedups. 

\end{appendix}

\newpage


\end{document}